\documentclass[aps,notitlepage,groupedaddress,superscriptaddress,11pt]{revtex4-1}

\usepackage{natbib}
\usepackage{pdfpages}
\pdfoutput=1

\makeatletter
\AtBeginDocument{\let\LS@rot\@undefined}
\makeatother

\usepackage{float}
\usepackage{graphicx}  
\usepackage{xcolor}
\usepackage{dcolumn}   
\usepackage{bm}        
\usepackage{amssymb}   
\usepackage{epstopdf}
\usepackage{xfrac}
\usepackage{subfigure}
\usepackage{anyfontsize}
\usepackage[export]{adjustbox}

\usepackage{amsmath}
\usepackage{amsfonts}
\usepackage{bbm}
\definecolor{darkblue}{rgb}{0.1,0.2,0.6}
\definecolor{darkred}{rgb}{0.8,0.1,0.2}
\usepackage[colorlinks,citecolor=darkblue,linkcolor=darkblue,urlcolor=darkblue]{hyperref} 
\usepackage{tikz}
\usepackage{enumerate}
\usepackage{setspace}
\usepackage{url}  
\usepackage{braket}
\usepackage{gensymb} 

\newcommand{\rv}{\textbf{r}}
\newcommand{\kv}{\textbf{k}}

\newcommand{\beginsupplement}{%
        \setcounter{table}{0}
        \renewcommand{\thetable}{S\arabic{table}}%
        \setcounter{figure}{0}
        \renewcommand{\thefigure}{{\bf S}\arabic{figure}}%
     }

\usepackage{fancyhdr}

\begin{document}

\clearpage
\onecolumngrid


# Quantum-Hall to Insulator Transition in Ultra-low-carrier-density Topological Insulator Films and a Hidden Phase of the Zeroth Landau Level

*Maryam Salehi[†], Hassan Shapourian[†], Ilan Thomas Rosen, Myung-Geun Han, Jisoo Moon, Pavel Shibayev, Deepti Jain, David Goldhaber-Gordon*, and *Seongshik Oh[*]*

Dr. M. Salehi
Department of Materials Science and Engineering, Rutgers, The State University of New Jersey, Piscataway, New Jersey 08854, USA

Dr. H. Shapourian
James Franck Institute and Kadanoff Center for Theoretical Physics, University of Chicago, IL 60637, USA

I. T. Rosen
Department of Applied Physics, Stanford University, Stanford, CA, 94305, USA

Dr. M.-G. Han
Condensed Matter Physics and Materials Science, Brookhaven National Lab, Upton, New York 11973, USA

Dr. J. Moon, P. Shibayev, D. Jain, and Prof. S. Oh
Department of Physics and Astronomy, Rutgers, The State University of New Jersey, Piscataway, New Jersey 08854, USA
E-mail: ohsean@physics.rutgers.edu

Prof. D. Goldhaber-Gordon
Department of Physics, Stanford University, Stanford, CA, 94305, USA.

Prof. D. Goldhaber-Gordon and I. T. Rosen
Stanford Institute for Materials and Energy Sciences, SLAC National Accelerator Laboratory, Menlo Park, CA, 94025, USA

arXiv:1903.00489v2 [cond-mat.mtrl-sci] 5 Jul 2019

**A key feature of the topological surface state under a magnetic field is the presence of the zeroth Landau level at the zero energy. Nonetheless, it has been challenging to probe the zeroth Landau level due to large electron-hole puddles smearing its energy landscape. Here, by developing ultra-low-carrier density topological insulator $Sb_2Te_3$ films, we were able to reach an extreme quantum limit of the topological surface state and uncover a hidden phase at the zeroth Landau level. First, we discovered an unexpected quantum-Hall-to-insulator-transition near the zeroth Landau level. Then, through a detailed scaling analysis, we found that this quantum-Hall-to-insulator-transition belongs to a new universality class, implying that the insulating phase discovered here has a fundamentally different origin from those in non-topological systems.**

A peculiar feature of topologically-protected Dirac surface states in topological insulators (TIs) is the existence of two-fold degenerate zeroth Landau levels (ZLL).[1-3] In the presence of structural symmetry (crystal inversion and top/bottom symmetry) and in the absence of tunneling between top and bottom surface states, the ZLL degeneracy is robust and cannot be lifted by the Zeeman energy (indeed they are merely shifted by Zeeman field).[4] Neglecting interactions, increasing the magnetic field such that the filling fraction approaches $\nu = 0$ causes the ZLLs to be half-filled, producing a compressible (metallic) state. This result holds even in the presence of non-magnetic random disorder which on average preserves the structural symmetry. In contrast, conventional two-dimensional electron gases (2DEGs) are insulating near $\nu = 0$, i.e., near the bottom of the lowest LL, due to the presence of localized states at the tail of the lowest LL.

Studying the physics of ZLL in TIs requires the Fermi level ($E_F$) to be close to the Dirac point. Access to this regime has proven challenging because high defect densities in TI systems push $E_F$ away from the Dirac point and also introduce significant level of electron-hole puddles

obscuring the energy landscapes of ZLL. Though LLs were seen with scanning tunneling spectroscopy,[5, 6] the quantum Hall effect (QHE) was not seen in transport in the early TI systems. One effective tool to suppress carrier density in V-VI TI systems is using a ternary or quaternary alloy(using a Bi and Sb mixture and/or a Se and Te mixture),[7, 8] lowering $E_F$ to where electrostatic-gating can tune the chemical potential to the Dirac point. With this method, the QHE was recently observed in transport on TI systems[9-11] but the properties of ZLL still remain elusive, presumably due to non-negligible electron-hole puddles.

In TI thin-films, the interface between the film and the substrate harbors significant concentration of defects, which push $E_F$ away from the Dirac point. A chemically/structurally-compatible substrate could help solve this problem, but no such substrate exists commercially. As an alternative, we have recently shown in $Bi_2Se_3$ thin-films that an optimally-designed buffer-layer suppresses the interfacial defects by more than an order of magnitude. These interface-engineered $Bi_2Se_3$ thin-films have led to a number of topological quantum effects including QHE in ungated TIs and quantized Faraday/Kerr rotations.[12-14] However, in the $Bi_2Se_3$ system, the ZLL is not clearly accessible due to proximity of the bulk valence band to the Dirac point.[15]

In this work, we have developed TI $Sb_2Te_3$ thin-films with carrier densities ($n_{sheet}$) as low as $1.0 \times 10^{11}$ cm$^{-2}$, the lowest reported in any TI system. As-grown $Sb_2Te_3$ films on sapphire harbor high density of interfacial defects and typically show a p-type sheet carrier density of ~$4.0 \times 10^{13}$ cm$^{-2}$. Grown on the optimally-designed buffer-layer (**Figure 1**), their carrier density immediately dropped to $n_{sheet} \approx 7 \times 10^{12}$ cm$^{-2}$ (p-type). When we introduced Ti as an n-type compensation dopant ($Ti^{4+}$ replacing $Sb^{3+}$), $n_{sheet}$ further dropped to as low as p-type of $1.0 \times 10^{11}$ cm$^{-2}$ and n-type of $1.4 \times 10^{11}$ cm$^{-2}$. These are over ten times lower than the previous best values observed in $(Bi_{1-x}Sb_x)_2Te_3$ thin-films,[16] which showed p-type of $1.4 \times 10^{12}$ cm$^{-2}$ and n-type of $1.5 \times 10^{12}$ cm$^{-2}$,

and also much lower than the best $Bi_2Se_3$ thin-films grown on optimal buffer layers,[15] which showed p-type of $2 \times 10^{12}$ cm$^{-2}$ and n-type of $6 \times 10^{11}$ cm$^{-2}$. Furthermore, Ti-doping to convert $Sb_2Te_3$ thin-films all the way into n-type is on its own surprising considering that all such attempts have previously failed for bulk crystals of $Sb_2Te_3$.[17-19] All the films were *in situ* capped with several layers of $(Sb_{0.65}In_{0.35})_2Te_3$ to protect against aging and preserve symmetry between the top and bottom surfaces of the TI film.

Figure 1a shows a schematic of the film structure on the $In_2Se_3/(Sb_{0.65}In_{0.35})_2Te_3$ buffer layer (BL). See Methods and Supporting Information for details of the film and BL growth. The first panel in Figure 1b shows high-angle annular dark-field scanning transmission electron microscopy (HAADF-STEM) of a 20QL-$In_2Se_3$/5QL-$(Sb_{0.65}In_{0.35})_2Te_3$/5QL-$Sb_2Te_3$/5QL-$(Sb_{0.65}In_{0.35})_2Te_3$ film (sections of this heterostructure are indicated by yellow dashed lines). The following four panels show false-color elemental mappings from energy-dispersive X-ray spectroscopy (EDS), demonstrating the presence/absence of In, Sb, Te, and Se elements in each layer.

**Figure 2** shows the transport properties of ungated BL-based 8QL-thick $Sb_2Te_3$ samples having Ti counter-dopant concentrations of 1.0% (ultra-low p-type; for ease we name it P8-1.0% where P refers to p-type sample, 8 is the thickness in quintuple layers, and 1.0% is the Ti concentration), 1.1% (p-type but closer to the n-p mixed regime (near the Dirac point); named P8-1.1%), 1.3% (very low n-type; named N8-1.3%), and finally 1.5% (ultra-low n-type; named N8-1.5%). See Figure S2 of Supporting Information for further transport properties of these films. Figure 2a and b show $\rho_{xx}$ and $\rho_{xy}$, respectively, at magnetic fields from 0 to 45 T at $T = 300$ mK. $v = 1$ QH plateaus are seen in both n- and p-type samples. In contrast, $Bi_2Se_3$ exhibited QHE only in n-type films.[5, 15, 20] Furthermore, the magnetic fields required to reach the QH state were as low

as 5.5 T for n-type (N8-1.5%) and 6.0 T for p-type films (see Figure S3 Supporting Information). This is at least 4 times smaller than that required for $Bi_2Se_3$. Remarkably, at higher fields, we observed a quantum-Hall-to-insulator transition (QIT) with the magneto-resistance ratio, defined as MR= $\frac{\rho_{insul-phase}(45T)-\rho_{QH-phase}}{\rho_{QH-phase}}$, where $\rho_{QH-phase}$ is the minimum $\rho_{xx}$ in the QH-regime, as large as $5 \times 10^3$ and $8 \times 10^4$ at $T = 300$ mK for N8-1.3% and N8-1.5%, respectively.

The transitions to QH and then to insulator show up more clearly in **Figure 3**a, the parametric flow lines of the conductivity tensor ($\sigma_{xx}$ versus $\sigma_{xy}$, where $\sigma_{xx} = \frac{\rho xx}{\rho_{xx}^2 + \rho_{xy}^2}$ and $\sigma_{xy} = \frac{\rho_{xy}}{\rho_{xx}^2 + \rho_{xy}^2}$) for sample N8-1.5% as the applied field is tuned over $0 < |B| < 35$ T. The trajectory includes a number of features: 1. Two outside curves with a cusp at low fields (from weak anti-localization), which approach $v = \pm 1$ plateaus ($\pm \frac{e^2}{h}$, 0) with increasing field as the sample enters the QH regime. 2. At higher fields, semicircles corresponding to the transition between the $v = \pm 1$ QH plateau at ($\pm \frac{e^2}{h}$, 0) and the insulating state at (0, 0). The insulating phase at $|B| > 24$T is characterized by a diverging $\rho_{xx}$ and a finite $\rho_{xy}$ (after anti-symmetrization) upon increasing field and decreasing temperature[21] (see Supporting Information for further discussion of this effect). The points ($\pm 0.5 \frac{e^2}{h}, 0.5 \frac{e^2}{h}$) mark the topological transition from the $v = \pm 1$ QH state to the insulating phase.

We propose two possibilities for the insulating behavior at high fields (near $v = 0$): either a coherent excitonic superfluid state between the degenerate ZLLs,[22] or a Hall insulator state.[21, 23] Coulomb interactions are essential for realizing the former candidate phase, while the latter phase can be stabilized in thin-films by disorder in strong magnetic fields even in the absence of interaction. Regarding the first possibility, our samples are well within the range (thickness $< 3\ell_B$,

where the magnetic length $\ell_B \approx \frac{26nm}{\sqrt{B}}$ ) where the inter-surface Coulomb interaction (between top and bottom surfaces) dominates over intra-surface Coulomb interactions (on top or bottom surfaces). Based on Reference [22], we might expect a coherent superfluid phase with a charge gap of ~16 meV. However, there is a competing process at strong magnetic fields leading to an enhanced tunneling between the top and bottom surfaces, which splits the zeroth LLs, favoring a Hall insulating phase (See Figure S11 of Supporting Information for more details). We theoretically estimate an upper bound for the magnetic field-induced gap to be ~10 meV. Unlike 2DEGs, this gap is not produced by the Zeeman spin splitting but is due to the quadratic correction to the Dirac dispersion near $\Gamma$ point: see Figure 3b and c, where the energy spectra of a TI thin-film and a 2DEG in the absence of magnetic field and in the presence of magnetic field with and without the Zeeman effect are schematically presented. The energy levels are explicitly derived in Supporting Information.

To better understand the nature of quantum criticality near the QIT, we investigate the transition's critical exponents.[24-27] Near the critical magnetic field at zero temperature, the system is characterized by power law divergence of the localization length $\xi \propto |B - B_c|^{-\nu}$. At finite temperatures, a new length-scale $L_\varphi \propto T^{-\frac{1}{z}}$ emerges, where $z$ is the dynamical critical exponent. This length is associated with inelastic scattering. The resistivity $\rho_{xx}$ assumes a finite-size scaling form $\rho_{xx}(B,T) = \frac{h}{e^2} f_1(\frac{\xi}{L_\varphi})$. **Figure 4**a (top panel) shows $\rho_{xx}$ of the sample N8-1.5% as a function of high magnetic fields from 18 T to 45 T for different temperatures from 300 mK to 10 K. The resistivity curves at different temperatures pass through a critical magnetic field ($B_c$) of 23.9T with $\rho_{xx}$ ($B_c$) = 0.75 $\frac{h}{e^2}$ and the critical filling fraction $\nu_c = 0.25$ (See Figure S4 in Supporting Information). Such critical field, resistivity, and filling fraction are together a hallmark of a scale-

invariant critical point. From the scaling hypothesis mentioned above, we have $\rho_{xx}(B,T) \propto |B - B_c|T^{-\kappa}$ near $B_c$, where $\kappa = \frac{1}{vz}$. The collapse of data as a function of the scaling-invariant parameter $|B - B_c|T^{-\kappa}$ up to 10 K gives $\kappa = 0.20 \pm 0.02$ in Figure 4a (bottom panel). The same $\kappa$ is extracted for N8-1.3%, indicating the universality of the scaling behavior (Figure S4 Supporting Information).

To evaluate $v$ and $z$ individually, however, we need to probe the dynamics of the system, which is done using the differential resistivity measurement. Like finite temperature, finite voltage (equivalently, electric field $E$) induces another length-scale $L_E \propto E^{-\frac{1}{1+z}}$ and the differential resistivity $\tilde{\rho}_{xx} = \frac{dV}{dI}$ takes the scaling form $\tilde{\rho}_{xx}(B,E) = \frac{h}{e^2} f_2(\frac{\xi}{L_E})$, implying that $\tilde{\rho}_{xx}(B,E) \propto |B - B_c|V^{-\kappa'}$ near $B_c$ where $\kappa' = \frac{1}{(z+1)v}$. The differential resistivity evolves from $\frac{d^2|V|}{dI^2} > 0$ in the QH state, to $\frac{d^2|V|}{dI^2} = 0$ at $B = B_c$, to $\frac{d^2|V|}{dI^2} < 0$ in the insulating state (Figure 4b, top panel). The critical scaling (Figure 4b, bottom panel) yields $\kappa' = 0.14 \pm 0.01$. Using the values of $\kappa'$ and $\kappa$, we extract $z = 2.1 \pm 0.2$ and $v = 2.4 \pm 0.3$. Additionally, we measured $\rho_{xx}$ as a function of magnetic field for different currents at a constant temperature (Figure 4c), again finding a critical point at $B_c = 24.2$ T. Similar exponents z = $1.9 \pm 0.3$ and $v = 2.7 \pm 0.5$ are extracted from the current scaling plot[28] in Figure 4c (See Figure S5 Supporting Information for detailed calculations).

The scaling behavior of QITs[27, 29-31] and plateau-to-plateau transitions[28, 32] in conventional 2DEGs always lead to $z \approx 1$. Although there is no fundamental reason why QITs should belong to the same universality class, there have been so far no exceptions in terms of the dynamical exponent. Our observation of $z \approx 2$ suggests that QIT in TSS belongs to a different universality class than QITs in conventional 2DEGs. The value $z = 1$ in 2DEGs is generally

attributed to unscreened Coulomb interactions[33, 34] and so our observation of $z = 2$ could indicate enhanced screening. We should note that $z = 2$ is the more naturally expected dynamical critical exponent for phase transitions in 2D instead of $z = 1$. An easy way to see this is via dimensional analysis. In general, a phase transition in a d-dimensional system which is characterized by a finite density of states yields $z = d$. After all, whether the difference in screening results from the different band structures (linear in TIs versus quadratic in 2DEGs) or excitonic superfluid interactions or other higher order effects is an open question for future studies. Further studies of QITs, specifically extracting $z$ and $\nu$, in other Dirac systems such as graphene could help resolve this question.

While the above QITs are driven by increasing magnetic field at fixed (small) carrier densities, **Figure 5** presents QITs driven by decreasing carrier density at fixed (large) magnetic fields. The sample is a top-gated, 8QL-thick $Sb_2Te_3$ film. By applying a positive gate voltage, we tuned the majority carrier from p- to n-type, realizing a well-defined QHE on the p-side and an incipient QHE on the n-side. Figure 5a and b show $\rho_{xx}$ and $\rho_{xy}$, respectively, as a function of gate-voltage at different magnetic fields: $\rho_{xy}$ is anti-symmetrized to eliminate mixing with large longitudinal resistance as explained in Supporting Information. Other details such as $\sigma_{xx}$ and $\sigma_{xy}$ as a function of $V_g$ and $n_{sheet}$, as well as the relationship between gate-voltage and $n_{sheet}$, can be also found in Supporting Information (Figure S6 and Figure S10). Figure 5c shows the parametric conductivity as the carrier density is tuned by gating: each line represents a gate sweep at a constant magnetic field. At high fields, the parametric conductivity follows semicircles with radius $0.5\frac{e^2}{h}$, indicating the transition from the $\nu = 1$ QH state to an insulating state, just as in Figure 3a. However, as the carrier density changes sign through the charge neutrality point (CNP), the flow

line exits the insulating phase and moves toward the opposite quantum Hall phase, which never occurs in the magnetic field-driven QIT in Figure 3a.

The ultra-low-carrier-density $Sb_2Te_3$ films allowed us to explore the zeroth LL of TSS that has been previously inaccessible due to large electron-hole puddles. In particular, we showed that the zeroth LL harbors a new insulating phase, with a distinct dynamical critical exponent. This implies that the origin of this insulating phase is fundamentally different from that of non-topological systems. This study has not only opened a new door for zeroth LL physics but also for testing various other proposals, including surface exotic phases[22, 35] and topological quantum computing.[36] In addition, considering that this bismuth-free TI system precludes any possible incorporation of $Bi_2Te_3$, whose Dirac point is buried in the valence band, it provides a promising new platform for high-temperature quantum anomalous Hall and other magneto-topological effects,[37-40] when combined with proper magnetic coupling schemes.

**Experimental Section**

*Growth:* All films were grown on 10 mm × 10 mm sapphire ($Al_2O_3$) (0001) substrates using a custom built SVTA MOS-V-2 MBE system with base pressure of low $10^{-10}$ Torr. After cleaning the substrates *ex situ* by a 5-minute exposure to UV-generated ozone, the substrates were heated *in situ* to 750 °C under oxygen pressure of $1 \times 10^{-6}$ Torr and were kept at 750 °C for ten minutes. The substrate was then cooled down to 170 °C for deposition of 3 quintuple layer (QL, 1QL ≈ 1 nm) $Bi_2Se_3$ which serves as a template for the following layer of $In_2Se_3$ deposited at 300 °C. The whole $Bi_2Se_3/In_2Se_3$ heterostructure was then heated to 600 °C so that $Bi_2Se_3$ diffused out past the $In_2Se_3$ layer and evaporated, leaving behind a highly insulating $In_2Se_3$ layer. To obtain ultra-low defect density $Sb_2Te_3$ thin-films, we next grew a more closely lattice-matched buffer – a 15QL-thick $Sb_{0.65}In_{0.35}Te_3$ layer, a solid solution of $Sb_2Te_3$ and $In_2Te_3$ – at 260 °C. This follows a strategy we previously developed for growth of low defect density $Bi_2Se_3$. After depositing the $Sb_2Te_3$ layer at 260 °C, we capped the structure with another 15QL-thick $Sb_{0.65}In_{0.35}Te_3$ at the same temperature for protection against air and preserving the symmetry of top and bottom surfaces.

For film growth, 99.999% pure elemental bismuth, indium, selenium, antimony, tellurium, and titanium (for doping) sources were thermally evaporated using Knudsen cells. Source fluxes were calibrated *in situ* with a quartz crystal micro-balance (QCM). The ratio of selenium flux to bismuth/indium flux and tellurium flux to antimony flux were maintained at above 10:1 and 5:1, respectively, as determined by QCM.

*Device fabrication:* Top-gated devices were fabricated using contact photolithographic patterning. For each patterning step, a hexamethyldisilazane adhesion layer was spin coated, followed by Megaposit SPR 3612 photoresist. The pre-exposure bake of 80 ˚C for 180 s was chosen to avoid thermal damage to the film. The photoresist was exposed under an ultraviolet mercury vapor lamp

at approximately 50 mJcm$^{-2}$ and was developed in Microposit developer CD-30 for 35 s. The device mesas were defined by photolithographically-patterned etching of the surrounding film with Ar ion milling. Ohmic contacts were made by first cleaning the area with an *in situ* Ar ion source, and then evaporating 5 nm Ti and 100 nm Au, followed by liftoff. To realize a robust top gate, a dielectric was grown uniformly across the film by evaporating a 1 nm Al seed layer, which was allowed to oxidize, and then depositing approximately 40 nm of alumina by atomic layer deposition. The top gate was then fabricated by evaporating 5 nm Ti and 85 nm Au, followed by liftoff. Excess alumina dielectric on the surrounding area was etched using Microposit developer CD-26 (tetramethylammonium hydroxide based, metal ion free). Metal was evaporated using a Kurt J. Lesker electron beam evaporator with an *in situ* Ar ion source. Atomic layer deposition used trimethylaluminum precursor and water as the oxidizer in a nitrogen purged vacuum chamber.

*Transport:* Magneto-resistance and Hall resistance measurements for the ungated samples were carried out using manually pressed indium leads in van der Pauw geometry as well as hand-scratched Hall bar geometry. For our transport measurements, we used various magnet systems including an electromagnet with magnetic field up to 0.6 T and a liquid He-4 cryogenic system with magnetic field up to 9 T in Rutgers University, as well as 35 T resistive and 45 T hybrid magnets at National High Magnetic Field Laboratory (NHMFL) in Tallahassee. To operate the 45 T hybrid magnet at NHMFL, the superconducting magnet is initially ramped up from 0 to 11 T and the rest of the field, 11 T to 45 T, is provided by the resistive magnet. Also, for the 45 T magnet, only one magnet polarization can be applied during a given measurement day, whereas sweeping through both positive and negative polarizations is possible in the 35 T resistive magnet.

Prior to measurement at NHMFL, the gated device was characterized at Stanford in a liquid He-4 cryogenic system (1.5 K base temperature, 14 T perpendicular magnetic field) as well as in

a He-3/He-4 dilution refrigerator (base temperature 20 mK, 14 T perpendicular magnetic field) using standard lock-in techniques. Due to the large resistances observed at high magnetic fields, quasi-DC techniques were used at NHMFL: the sample was biased with a DC current whose polarity alternated approximately every 1.5 s; the resistance was taken to be the anti-symmetric component between the value under positive and negative bias polarity. Anti-symmetrization in bias polarity removes spurious thermoelectric contributions from the wiring. All reported resistances are four-terminal voltage measurements divided by the constant current bias. The sheet carrier density was calculated from $R_{xy}$ using the Hall formula $n_{sheet} = \left(e \frac{dR_{xy}}{dB}\right)^{-1}$ where $\frac{dR_{xy}}{dB}$ was taken from the linear part of low field measurement and e is the electronic charge. The carrier mobility ($\mu$) was then calculated using $\mu = (e \, \rho_{xx} \, n_{sheet})^{-1}$, where $\rho_{xx}$ is the sheet resistance near zero magnetic field.

## Supporting Information

Supporting Information is available from the Wiley Online Library or from the author.

## Acknowledgements

[†]Maryam Salehi and Hassan Shapourian contributed equally to this work.

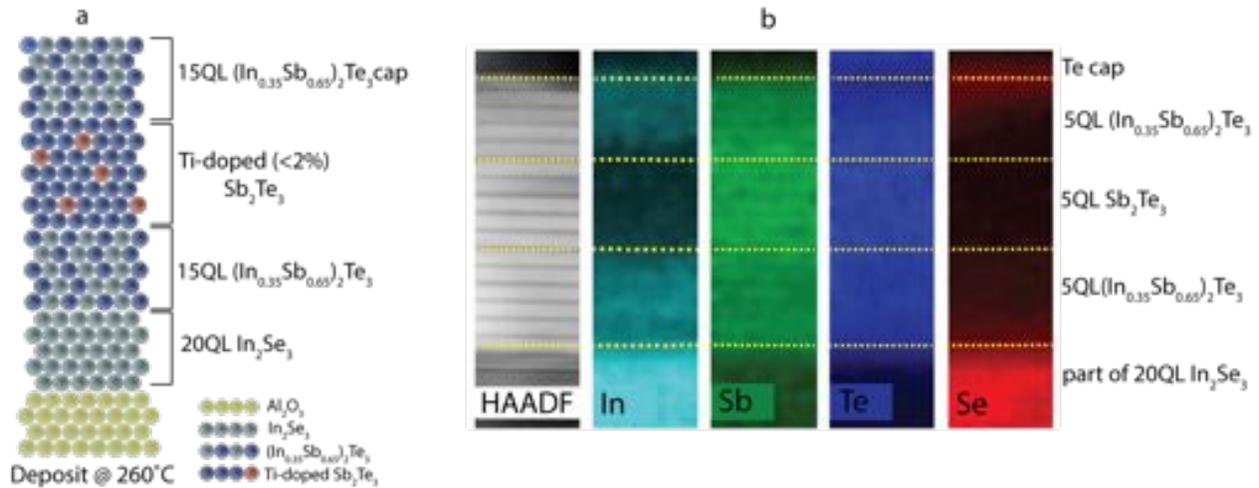

**Figure 1.** Structure of an ultra-low-carrier-density $Sb_2Te_3$ film. a) Schematic of the film structure which consists of 20QL $In_2Se_3$ grown on $Al_2O_3(0001)$ substrate, using a sacrificial seed layer of $Bi_2Se_3$, followed by growth of 15QL $(Sb_{0.65}In_{0.35})_2Te_3$ buffer layer as a template for the successive layer of ultra-low-carrier-density Ti-doped $Sb_2Te_3$. Finally, the whole film is capped by 15QL $(Sb_{0.65}In_{0.35})_2Te_3$ (see Reference [12], Method section, and Supporting Information for more detail on growth). b) The first panel shows high-angle annular dark-field scanning transmission electron microscopy (HAADF-STEM) for 20QL-$In_2Se_3$/5QL-$(Sb_{0.65}In_{0.35})_2Te_3$/5QL-$Sb_2Te_3$/5QL-$(Sb_{0.65}In_{0.35})_2Te_3$ with an additional 10 nm Te capping to protect the whole super-lattice against the STEM sample preparation processes. Yellow dashed lines drawn at each interface are guides to the eye. The rightward four panels show the elemental mapping energy-dispersive X-ray spectroscopy (EDS) images for In (light blue), Sb (green), Te (dark blue), and Se (red). If a specific element is present in a layer, then the corresponding color appears bright in that section; in its absence, the layer appears dark. For instance, $Sb_2Te_3$ contains Te and thus appears as bright blue in the fourth panel, but it lacks Se and thus appears as dark in the fifth panel.

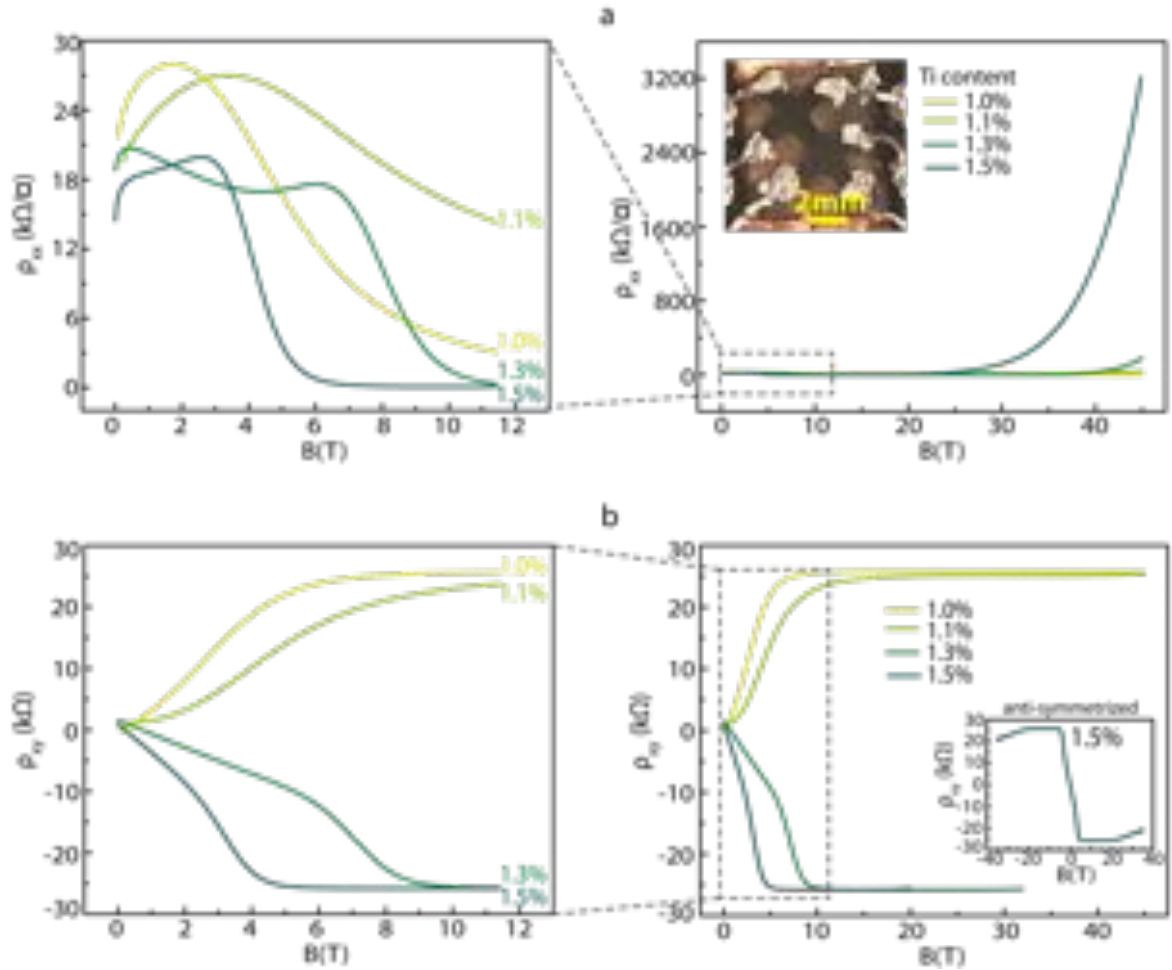

**Figure 2.** Transport Properties of 8QL films with different Ti-doping levels. a) $\rho_{xx}$ for 8QL $Sb_2Te_3$ films with 1.0%, 1.1%, 1.3% and 1.5% of Ti-doping as a function of magnetic field from 0 to 45 T. Inset shows a photo of the ungated device with a hand-scratched Hall bar and manually-pressed indium wires as contact leads. The left panel is for 0 < B < 11 T. b) $\rho_{xy}$ as a function of magnetic field from 0 to 45 T. The left panel is for 0 < B < 11 T. The inset in b (right panel) shows the anti-symmetrized Hall resistivity in N8-1.5%.

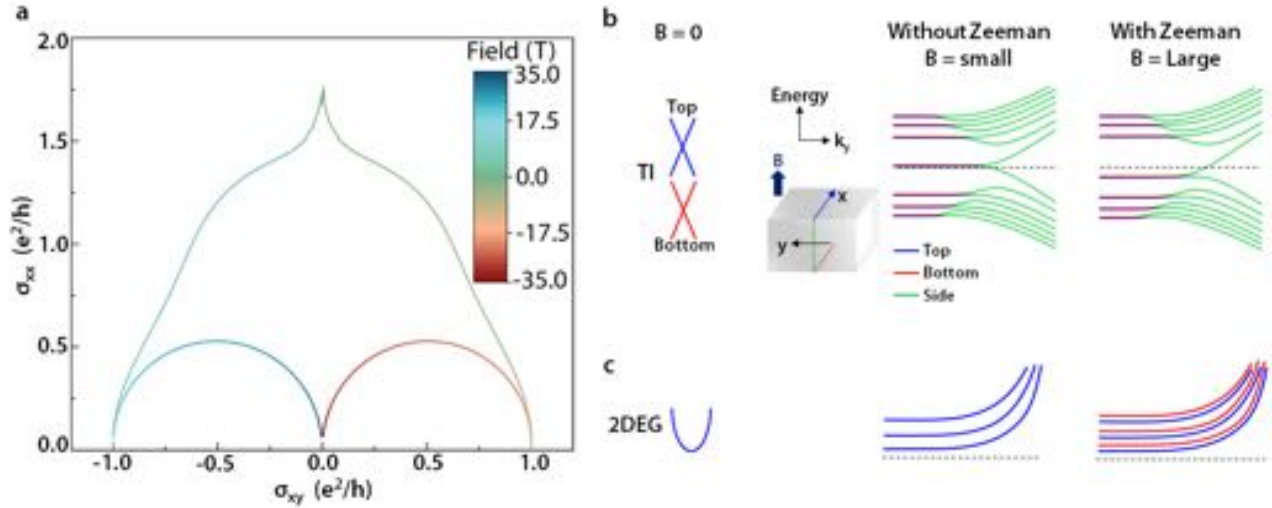

**Figure 3.** Conductivity tensor flow lines of an 8QL film and comparison of LLs along with edge modes in a magnetic field. a) The conductivity tensor flow ($\sigma_{xx}$ vs. $\sigma_{xy}$) of N8-1.5% for $|B| \leq 35$ T. The field is incorporated as a color map in the plot. The points ($\pm e^2/h$,0), (0,0), and ($\pm 0.5e^2/h$, $0.5e^2/h$) correspond to the QH state, the insulator phase, and the transition between these two phases, respectively. The cusp around zero field indicates weak anti-localization effect. Schematic of energy spectrum of b) TI thin-film c) 2DEG in the absence of magnetic field (B = 0; the first column), in a small magnetic field without the Zeeman-coupling (B = small; the second column) and in a strong magnetic field with the Zeeman-coupling (B = large; the third column). The zero energy is marked by dashed lines. Because of spin-momentum locking in TI surface Hamiltonian, the Zeeman field does not lift the LL degeneracies; instead, it shifts all LLs. Particularly, the Zeeman term moves the top and bottom LLs up or down depending on magnetic field direction. The energy shift is more pronounced for the zeroth LL and becomes smaller for higher LLs (See Supporting Information for more detail). In contrast, in 2DEG (also in graphene; see Figure S17 of Supporting Information) the Zeeman field lifts the spin degeneracy (here spin-split 2DEG is shown).

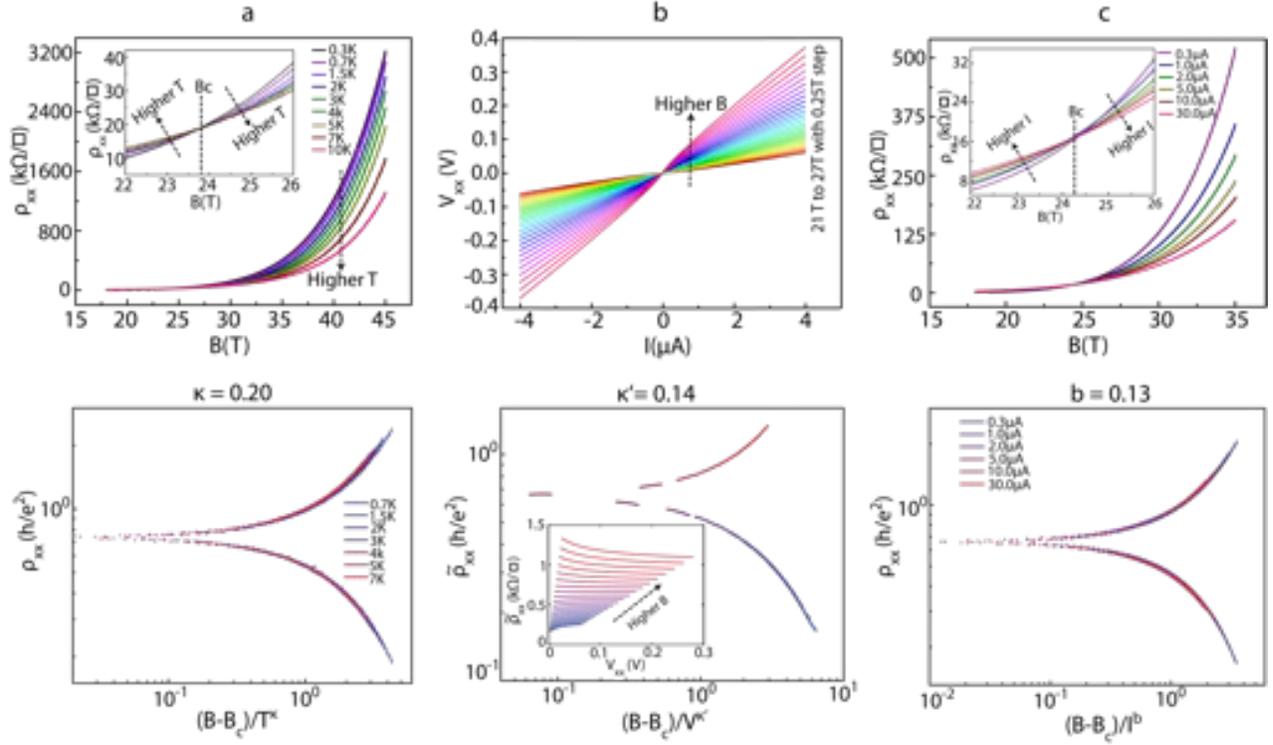

**Figure 4.** Scaling-invariant plots for an 8QL film. a) $\rho_{xx}$ as a function of magnetic field for N8-1.5% at temperatures from 300 mK to 10 K (See Supporting Information for the flow of conductivity tensor at different temperatures). The critical magnetic field ($B_c \approx 23.9$ T), marked by a dashed line in the inset, is where all the curves pass through. The bottom panel is the corresponding scale-invariant plot for temperatures, yielding $\kappa = 0.20 \pm 0.02$. b) Top panel shows I-V curves for the 1.5%Ti-doped sample at a constant field ranging from 21 T to 27 T with 0.25 T step. Bottom panel is the corresponding scale invariant plot and the inset shows $\widetilde{\rho_{xx}}$ as a function of $V_{xx}$. $\kappa' \approx 0.14$ is extracted from the fit. c) $\rho_{xx}$ as a function of magnetic field for the 1.5%Ti-doped sample for different currents. Like the temperature-dependent curves in (a), the curves corresponding to different currents pass through a critical magnetic field ~24.2 T (the small shift results from sample aging). The bottom panel shows the corresponding scale-invariant plot for current.

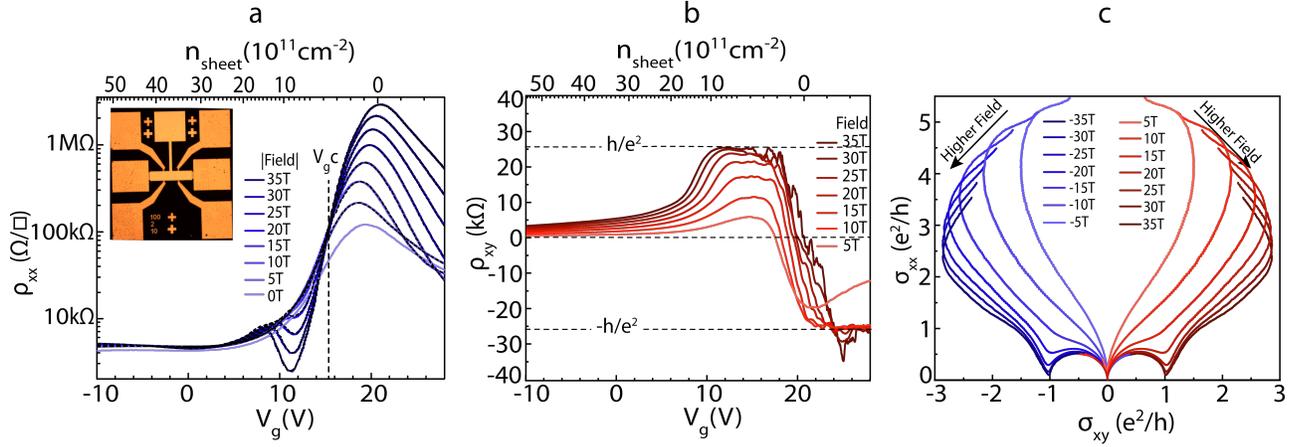

**Figure 5.** Transport properties of a gated 8QL film with a 2nm $(Sb_{0.65}In_{0.35})_2Te_3$ capping. a) $\rho_{xx}$ as a function of gate-voltage ($-10$ V $\leq V_g \leq 28$ V) for different fields 0 T, 5 T, 10 T, 15 T, 20 T, 25 T, 30 T, and 35 T at 300 mK. Although vanishing $\rho_{xx}$ was not observed, the QH regime is indicated by a dip near $V_g = 12$ V at B > 20 T, reaching 2.4 kΩ/sq at 35 T. The peak (~119 kΩ/sq at zero field and ~2.9 MΩ/sq at 35T) corresponds to the CNP. Curves at B > 10 T pass through a critical point $V_g^C = 15.4$ V. At higher gate-voltages, an incipient n-type QHE is observed. Inset shows a photo of the fabricated device. b) Anti-symmetrized $\rho_{xy}$ as a function of gating from $-10$ V to 28 V for the same fields. The $\pm h/e^2$ plateaus are marked with dashed lines on the plot. The noisiness at high fields comes from incomplete elimination of the geometrically-mixed large $\rho_{xx}$ component during the anti-symmetrization process. c) The parametric flow of the conductivity tensor ($\sigma_{xx}$ vs. $\sigma_{xy}$) after (anti-)symmetrization. Each line represents a gate sweep from $V_g = -10$ V to 28 V at a constant field.



\beginsupplement


\begin{center}
$ $
\vspace{2cm}

{\Large Supporting Information for}
\vspace{1cm}

{\bf\large Quantum-Hall to Insulator Transition in Ultra-low-carrier-density Topological insulator Films and a Hidden Phase of the Zeroth Landau Level}

Maryam Salehi,
Hassan Shapourian,
{Ilan Thomas Rosen}, 
{Myung-Geun Han},\\
{Jisoo Moon},
{Pavel Shibayev},
{Deepti Jain},
{David Goldhaber-Gordon},
{Seongshik Oh$^\ast$}

\vspace{0.3cm}
\centerline{$^\ast$Corresponding author. Email: \email{ohsean@physics.rutgers.edu}{ohsean@physics.rutgers.edu}}
 \end{center}


\vspace{3cm}








\section{Supplementary Figures (Experiment)}

\begin{figure}[H]
\centering
 \includegraphics[scale=0.11]{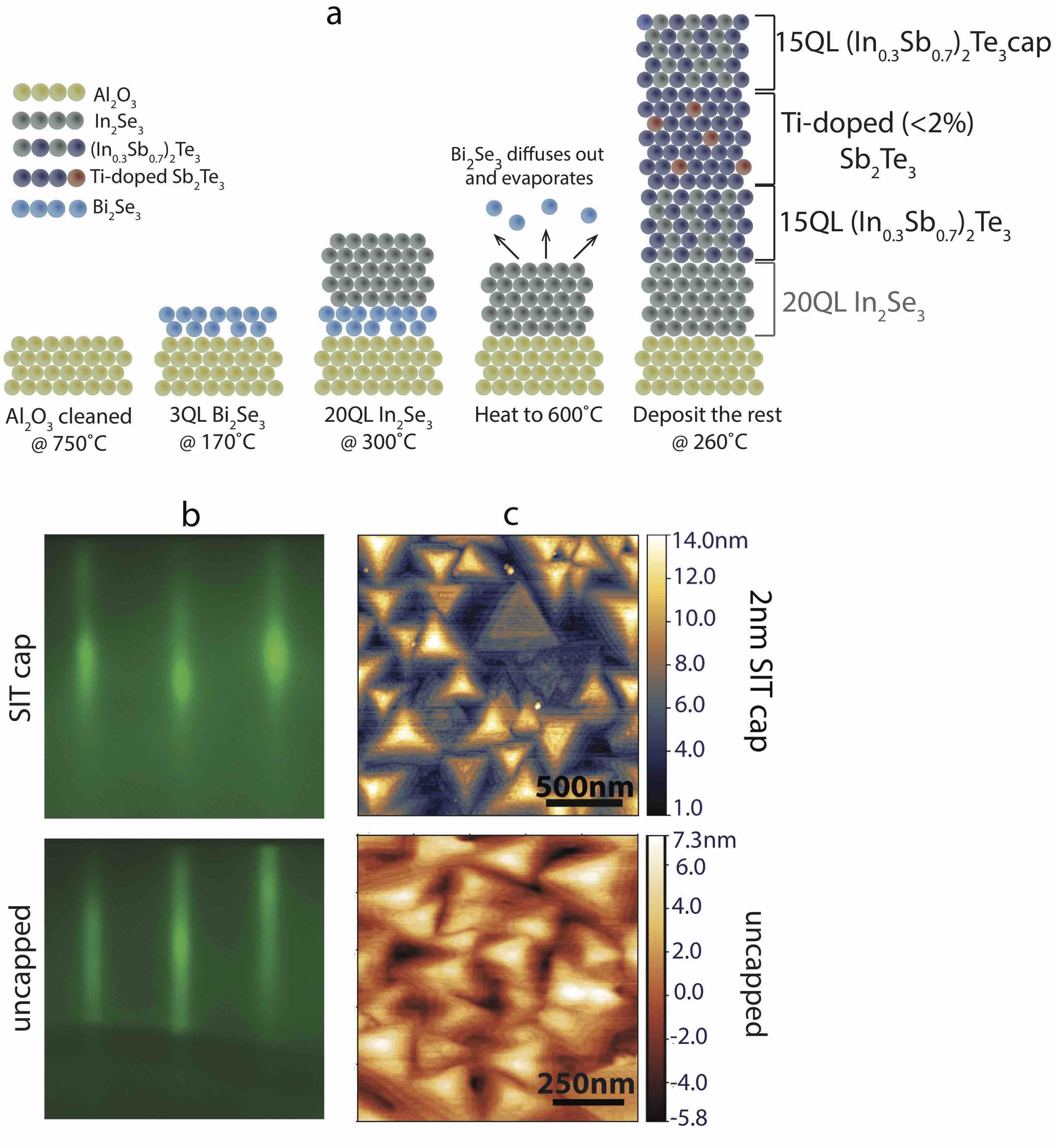}
\caption{\label{fig:growthprocedure}a) Schematic of the growth procedure of a buffer layer-based Sb$_2$Te$_3$ film. We chose (Sb$_{1-x}$In$_x$)$_2$Te$_3$ (SIT), a solid solution of trivial insulator  In$_2$Te$_3$ and topological insulator Sb$_2$Te$_3$, as the template for Sb$_2$Te$_3$. However, since In$_2$Te$_3$ has a different crystal structure (defect zinc blende lattice with \textit{a} = 6.15\AA) than Sb$_2$Te$_3$,  the growth structure degrades above a certain In concentration~\cite{ROSENBERG1961105}. Therefore, an optimized amount of In has to be added so that the structure remains crystalline and yet far from metallic Sb$_2$Te$_3$ and as insulating as possible. We noticed that above 40\% In, the structure starts to degrade with visible formation of 3D spots in the reflection high energy electron diffraction (RHEED) image. Therefore, due to sample to sample variation and to be safe, we decided to use (Sb$_{0.65}$In$_{0.35}$)$_2$Te$_3$ for both the buffer and capping layers in all samples to stay slightly lower than 40\% where the structure is on the verge of becoming bad. b) Streaky RHEED images for Sb$_2$Te$_3$ film grown on the SIT  buffer layer and before it gets capped (bottom panel) and for the capping layer which is grown at the same growth temperature (top panel) show a flat 2D growth of high quality film and capping. c) Atomic force microscopy (AFM) on both Sb$_2$Te$_3$ film with no capping (bottom panel) and on 2nm-thick (Sb$_{0.65}$In$_{0.35}$)$_2$Te$_3$ capping (top panel) shows triangular traces indicative of three-fold symmetry.  }
\end{figure}
\newpage

\begin{figure}[H]
\centering
\includegraphics[scale=0.1]{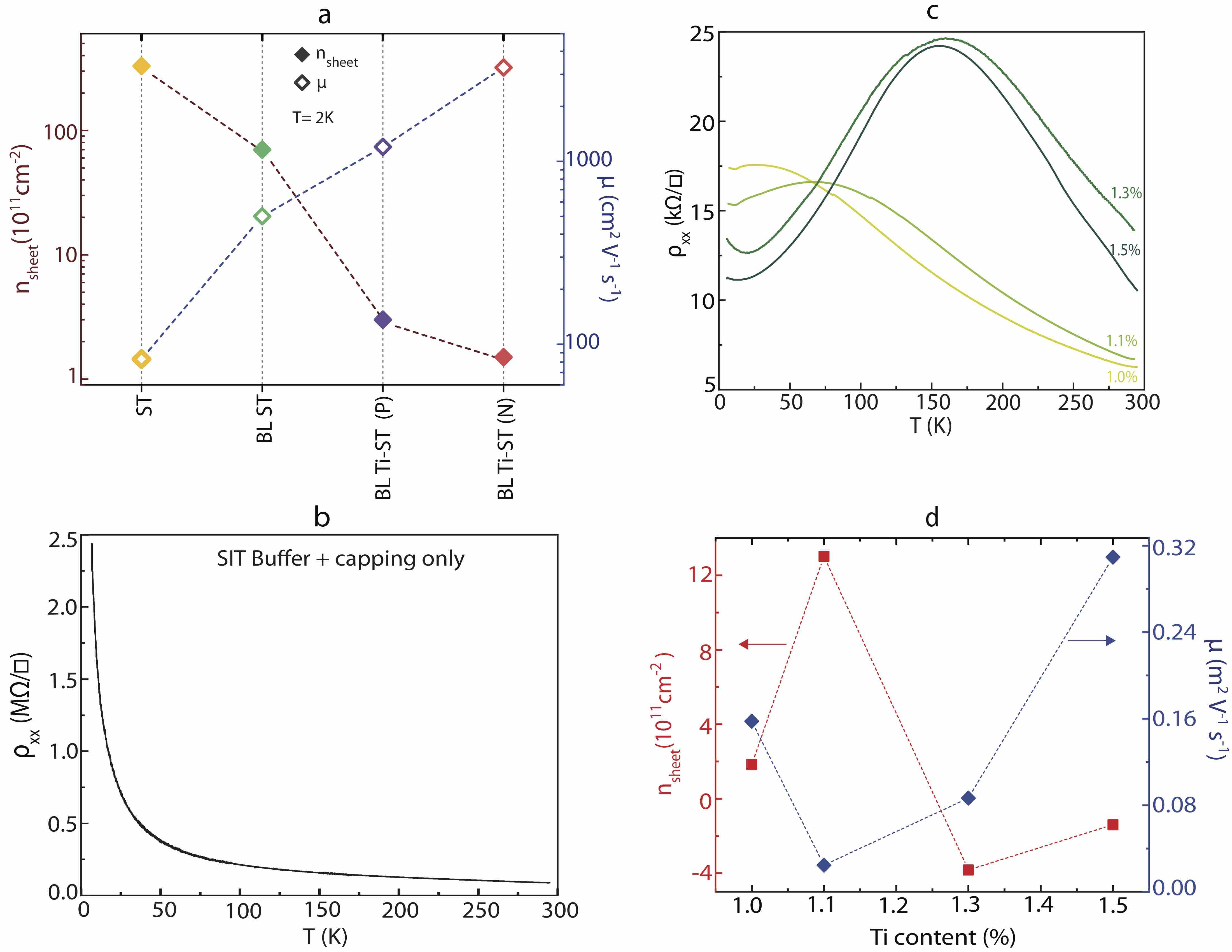}
\caption{\label{fig:transport}a) Comparison of the lowest achieved carrier densities (solid diamonds connected by a dark red dashed line as a guide to the eye) with corresponding mobilities on the right axis (hollow diamonds connected by a blue dashed line) for Sb$_2$Te$_3$ films grown directly on sapphire (ST), buffer layer(BL)-based Sb$_2$Te$_3$ (BL-ST), and Ti-doped SIT BL-based Sb$_2$Te$_3$ (p-type: BL Ti-ST (P) and n-type: BL Ti-ST (N)). b) Sheet resistance as a function of temperature for a 30QL (Sb$_{0.65}$In$_{0.35}$)$_2$Te$_3$ sample (buffer + capping only) has insulating temperature-dependent behavior where it is not fully insulating at room temperature and becomes insulating at low temperatures. It is worth noting that a thin layer ($\sim 10$ to $20$ nm) of Te capping which has been used in some of the previous studies~\cite{Zhang_bandstructure,Chang_Highprecision} is not completely insulating at room temperature either, but becomes highly resistive at low temperature. c) Sheet resistance as a function of temperature (from room temperature down to 6 K) for four Ti-doped samples: P8-1\%, P8-1.1\%, N8-1.3\%, and N8-1.5\% (mentioned in the main text as well). d) Transport properties (\textit{n$_{sheet}$} and $\mu$) of the same four samples.  By adding Ti, \textit{n$_{sheet}$} decreases to ultra-low p (1.8$\times$10$^{11}$ cm$^{-2}$) in sample P8-1\%. As more Ti is added and as the Fermi level gets closer to the Dirac point (mixed n- and p-zone), the \textit{n$_{sheet}$} artificially increases to 1.3$\times$10$^{12}$ cm$^{-2}$ in P8-1.1\% and eventually  additional Ti leads to an n-type sample, N8-1.3\% with n$_{sheet}$ = -3.8$\times$10$^{11}$ cm$^{-2}$, and even to a lower carrier density sample, N8-1.5\% with  n$_{sheet}$ = -1.4$\times$10$^{11}$ cm$^{-2}$. Adding Ti decreases the mobility value on the p-type side, but $\mu$ increases again on the n-type side. Beyond a certain point, adding more Ti makes the system insulating.}
\end{figure}
\begin{figure}[H]
\centering
\includegraphics[scale=0.1]{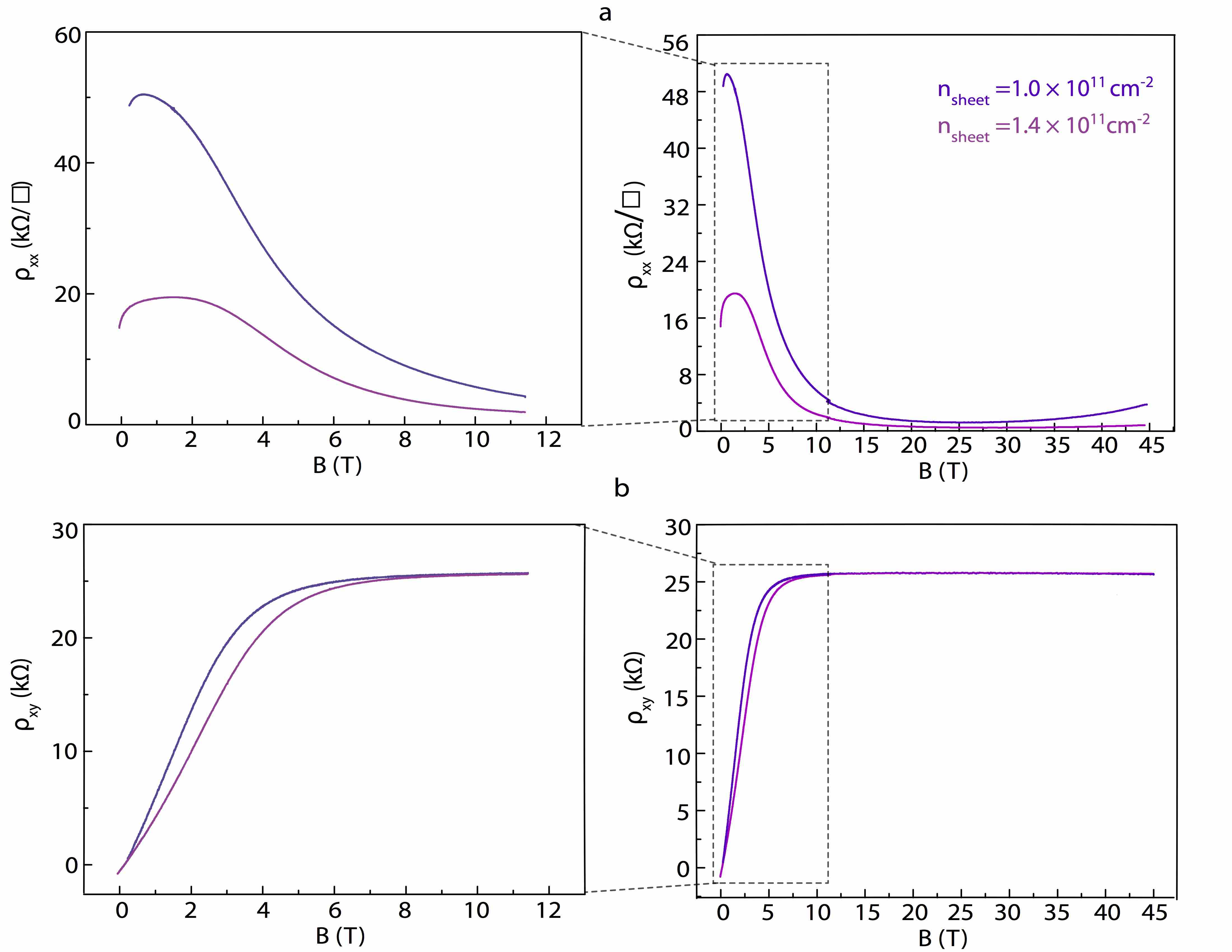}
\caption{\label{fig:6qltransport}a) $\rho_{xx}$  as a function of magnetic field (from 0 to 45 T; right panel) in two 6QL samples with slightly different carrier densities of 1.0$\times$10$^{11}$ cm$^{-2}$ (with mobility 1373 cm$^2$V$^{-1}$s$^{-1}$) and 1.4$\times$10$^{11}$ cm$^{-2}$ (with mobility 3065 cm$^2$V$^{-1}$s$^{-1}$)  at 300 mK (the difference comes from sample to sample variation). Left panel is a zoomed-in $\rho$$_{xx}$ plot from 0 to 11 T. b) $\rho$$_{xy}$ as a function of magnetic field (from 0 to 45 T) in the same samples. Left panel is a zoomed-in $\rho$$_{xy}$ plot from 0 to 11 T. This confirms that we observed the quantum Hall effect (QHE) at low magnetic fields for ungated p-type samples as well. Upon adding more Ti, the 6QL sample turns  insulating before becoming n-type.
}
\end{figure}
\begin{figure}[H]
\centering
\includegraphics[scale=0.1]{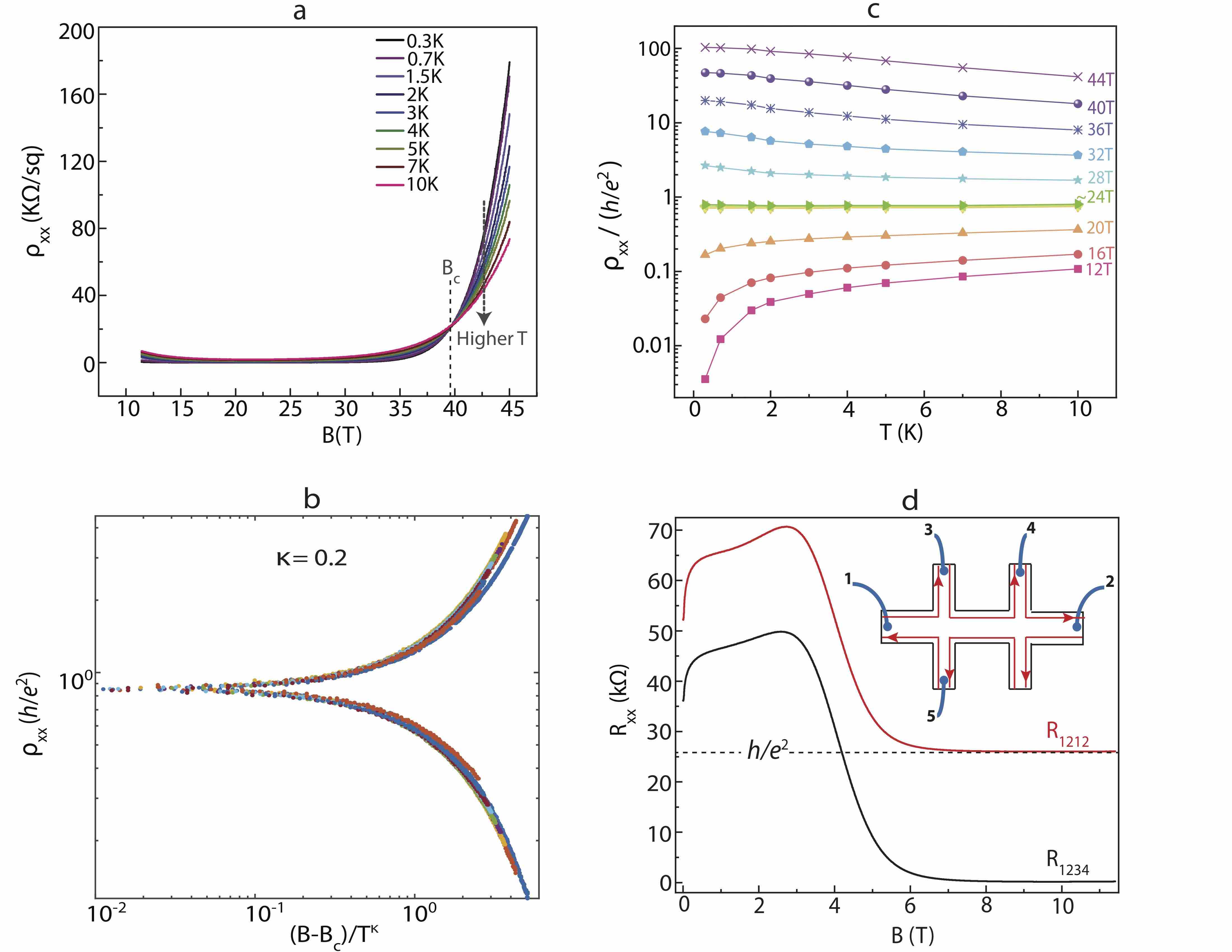}
\caption{\label{fig:otherscaling}a) $\rho$$_{xx}$ as a function of magnetic field for N8-1.3\% sample at different temperatures (300 mK to 10 K). The critical magnetic field ($B_c$) is 38.5 T (marked by a dashed line) where all the curves cross. b) The corresponding temperature scale-invariant plot yields $\kappa=0.2$ which is the same as the N8-1.5\% sample mentioned in the main text, showing the universality and sample-independence of the scaling behavior. c) Normalized sheet resistance (divided by the resistance quantum) as a function of temperature for N8-1.5\% sample. The $\rho_{xx}$ values for different temperatures (300 mK to 10 K) at a constant field are taken from Fig. 4{\bf a} in the main text. Below $B_c$ and at low temperatures, as the sample enters QH regime, $\rho_{xx}$ vanishes. At the critical point $B_c$ = 23.9 T, $\rho_{xx}$ should be constant and is $0.75h/e^2$. Above $B_c$, as the sample transitions to insulating phase, $\rho_{xx}$ grows to a large number. d) $R_{xx}$ of two-point ($R_{1212}$; running current through contacts 1 and 2 and measuring voltage between the same leads) and four-point  ($R_{1234}$; running current through contacts 1 and 2 and measuring voltage between leads 3 and 4) measurements for N8-1.5\%. $R_{1212}$ starts from $\sim$1.46$R$$_{1234}$ and eventually converges to $h/e^2$ (+250 $\Omega$ contact resistance) which confirms the bulk of the sample is insulating and the edge is conducting (perfect QHE). In contrast, $R_{1234}$ vanishes as the sample enters the QH regime.}
\end{figure}
\begin{figure}[H]
\centering
\includegraphics[scale=0.1]{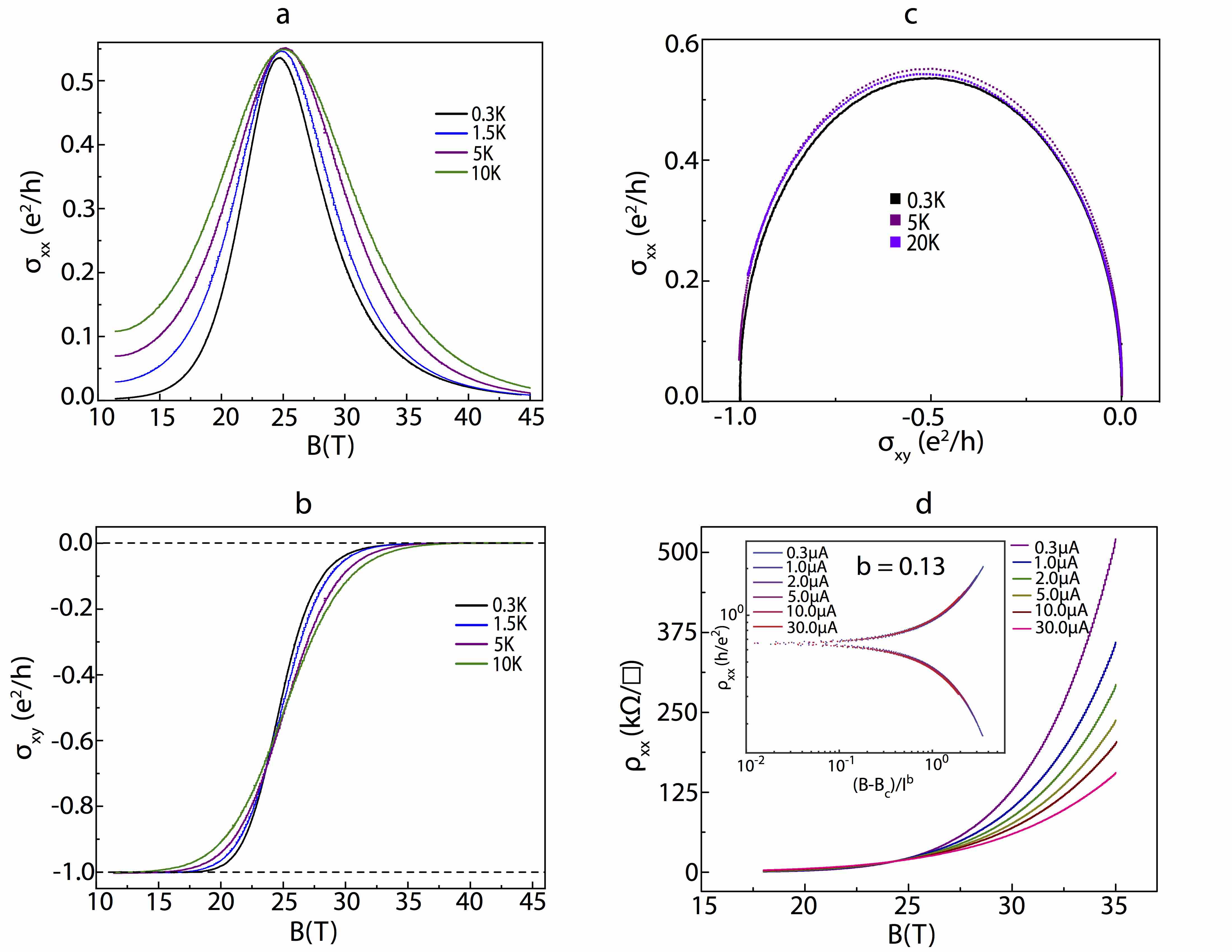}
\caption{\label{fig:scaling}  Magnetic field dependence of a) $\sigma$$_{xx}$ and b) $\sigma$$_{xy}$ of N8-1.5\% sample for diffrent temperatures.  $\sigma$$_{xx}$'s peak and $\sigma$$_{xy}$'s cross at the critical magnetic field ($B_c$). The $\nu=0$ and $\nu=-1$ plateaus are marked by dashed lines in the $\sigma$$_{xy}$ plot. c) The  flow lines of conductivity tensor ($\sigma$$_{xx}$ vs. $\sigma$$_{xy}$) in the same sample  where the results at 3 different temperatures and from 11 T to 45 T collapse on a semicircle-like trajectory extending from (-1,0), corresponding to the QH phase, to (0,0) for the insulating phase. (-0.5,0.5) represents the transition point between these two phases. The signature of quantum Hall-to-insulator transition can be observed even at higher temperatures (as high as 20 K). d) Magnetic field dependence of $\rho$$_{xx}$ of N8-1.5\% sample for different currents  (also shown in Fig. 4c of the main text). The curves corresponding to different currents cross at a critical magnetic field $\sim$24.2 T (the small shift of $B_c$ compared to the one in the temperature plot, Fig. 4a of the main text, could be due to sample aging). The inset shows the corresponding current scale-invariant plot. The underlying assumption is that the dissipated energy in the system effectively heats up the electrons via $ k_{B}T_e$ $\sim$ $eEL_{\phi}(T_e)$ which implies that the electron temperature $T_e$ is related to the current I as $T_e$ $\propto$ $I^{z/(1+z)}$. Therefore, the current-scaling plot admits the scaling form $\rho$$_{xx}$ $\propto$ $\left|{B - B_c}\right|$$I^{-b}$, where $b=1/\nu(1+z)$. Combining $b$ and $\kappa$$^\prime$, we found that $z= 1.9\pm 0.3$ and $\nu=2.7\pm0.5$. }
\end{figure}
\begin{figure}[H]
\centering
\includegraphics[scale=0.1]{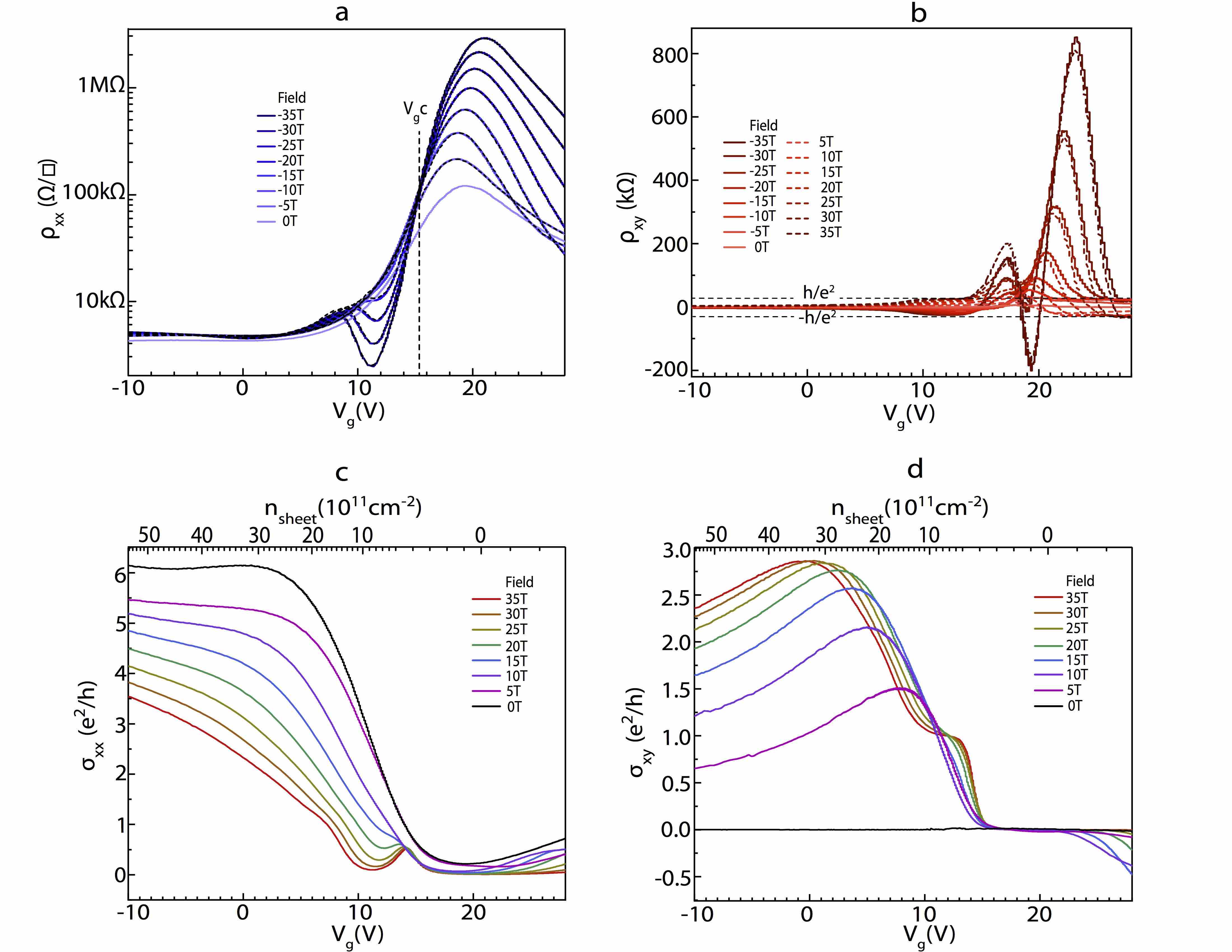}
\caption{\label{fig:R12gating}a) Raw data for $\rho$$_{xx}$ (Fig. 5a in the main text) as a function of gate-voltage (-10 V $\leq V_g \leq$ 28 V) for different fields of 0 T, 5 T , 10 T, 15 T, 20 T, 25 T, 30 T, and 35 T at 300 mK. b) Raw data for $\rho$$_{xy}$ as a function of gate-voltage (10 V $\leq V_g \leq$ 28 V) for the same fields with $h/e^2$ and  $-h/e^2$ plateaus corresponding to the QHE on p and n sides, respectively. The large peaks in $\rho_{xy}$ are due to mixing with large $\rho_{xx}$ around the charge neutrality point. The anti-symmetrized result for $\rho_{xy}$ is shown in Fig. 5b of the main text. c) $\sigma_{xx}$ as a function of gate-voltage 10 V $\leq V_g \leq$ 28 V (bottom axis) and sheet carrier density (top axis) for the different fields. d) $\sigma_{xy}$ as a function of  gate-voltage (10 V$\leq V_g \leq$ 28 V) and sheet carrier density for the same fields, where  $\nu$ = 0 and $\nu$ = 1 plateaus are visible. }  
\end{figure}

\begin{figure}[H]
\centering
\includegraphics[scale=0.1]{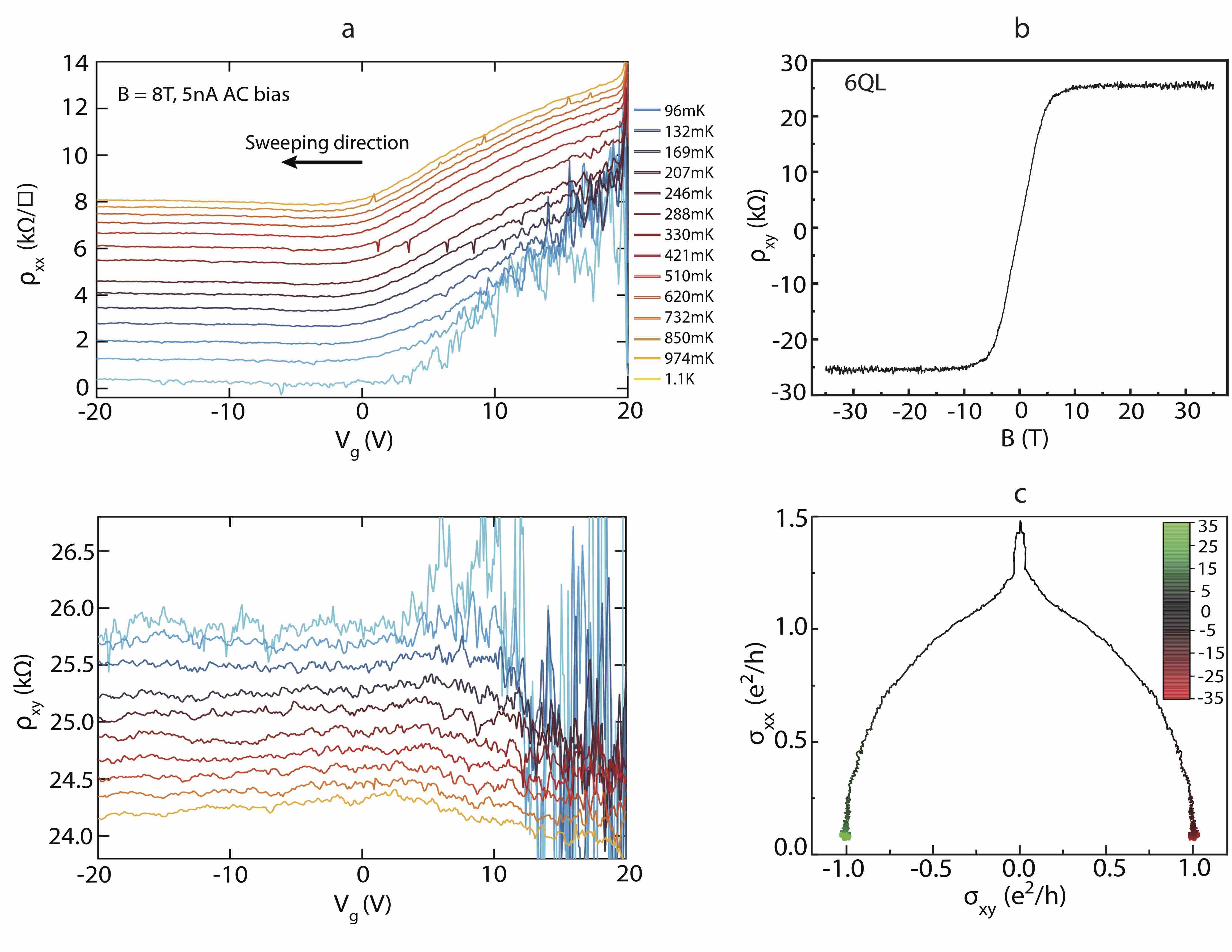}
\caption{\label{fig:gatedR4}a) $\rho$$_{xx}$ (top panel) and  $\rho$$_{xy}$ (bottom panel) at different temperatures for a 6QL sample with 0.7\% Ti-doping (P6-0.7\%). At ultra-low temperatures, a well-defined QHE with vanishing $\rho$$_{xx}$ is achievable, most likely due to suppression of thermally-activated dissipation possibly coming from buffer and capping layers. b) Anti-symmetrized result for the 6QL sample at 35 T. c) The flow of the same sample for magnetic fields from -35 T to 35 T. The field is incorporated as a color map in the plot. ($\pm e^2/h,0$) points corresponds to QH regime, (0,0) represents the insulating phase, and  ($\pm 0.5 e^2/h, 0.5 e^2/h$) corresponds to the transition between these two phases. The cusp at low fields indicates the weak anti-localization.}
\end{figure}

\begin{figure}[H]
\centering
\includegraphics[scale=0.49]{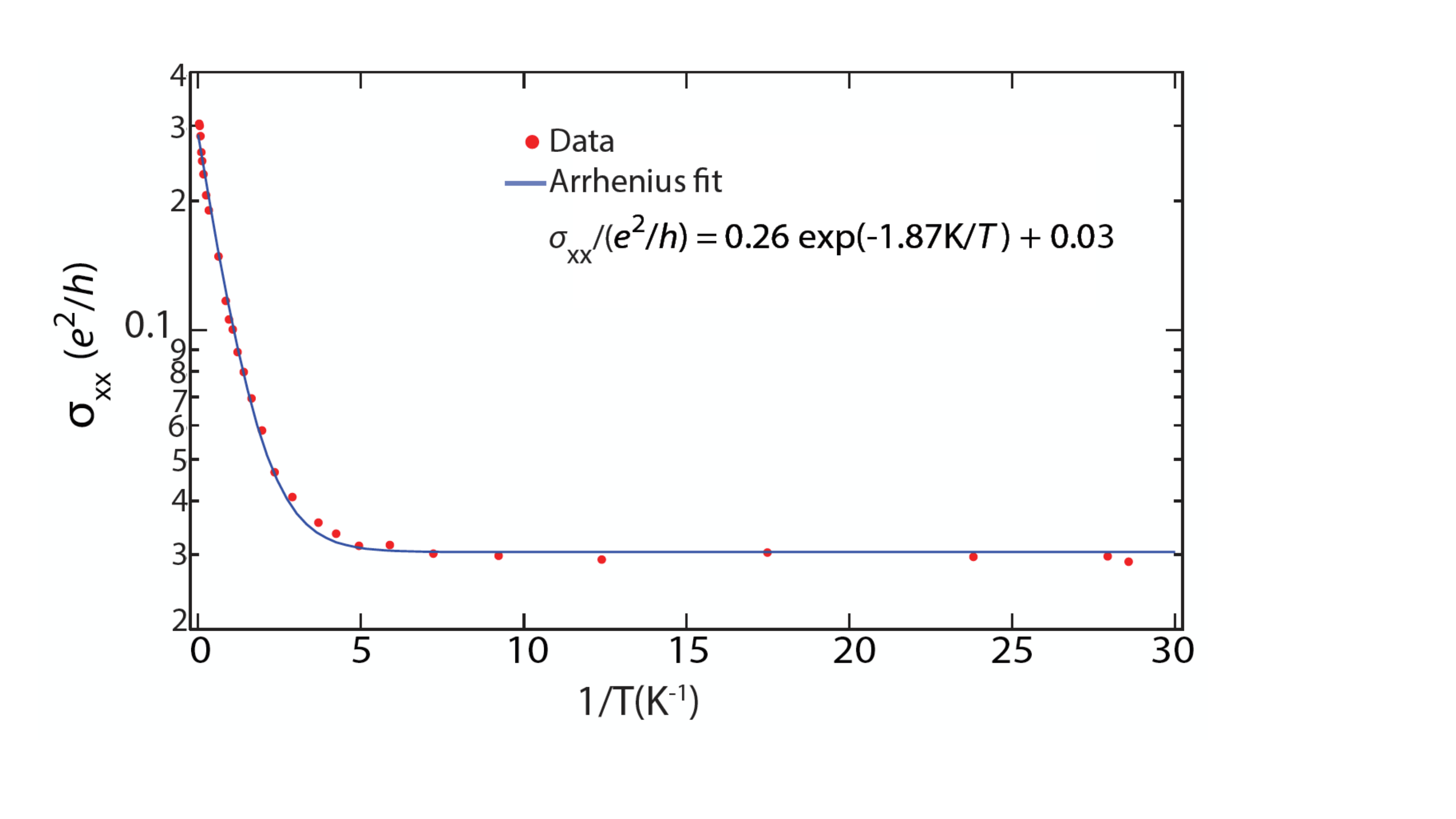}
\caption{\label{fig:sxx1overT}The longitudinal conductivity of the gated device at zero field, gate-tuned to the charge neutrality point, shown on a log scale as a function of inverse temperature. The data is fit to an Arrhenius model $\sigma_{xx}$ $\approx$ $\exp(-\Delta_t/k_B T)+\sigma^0_{xx}$, finding $\Delta_t$ = 161 $\mu\text{eV}$. The gap at charge neutrality is understood as a consequence of hybridization between the top and bottom surface states. We attribute the constant offset $\sigma^0_{xx}$  to Joule heating as its value was observed to decrease with decreasing current bias. This measurement used a 1 nA current bias for temperatures $T<1.5 \text{ K}$. This measurement included data from two separate cooldowns (one in a He-3/He-4 dilution refrigerator and one in a He-4 system). To account for offsets in gating between the two cooldowns, the conductivity shown is the minimum value of $\sigma_{xx}$ (as a function of gate-voltage) at each temperature.}
\end{figure}
\newpage
\begin{figure}[H]
\centering
\includegraphics[scale=0.75]{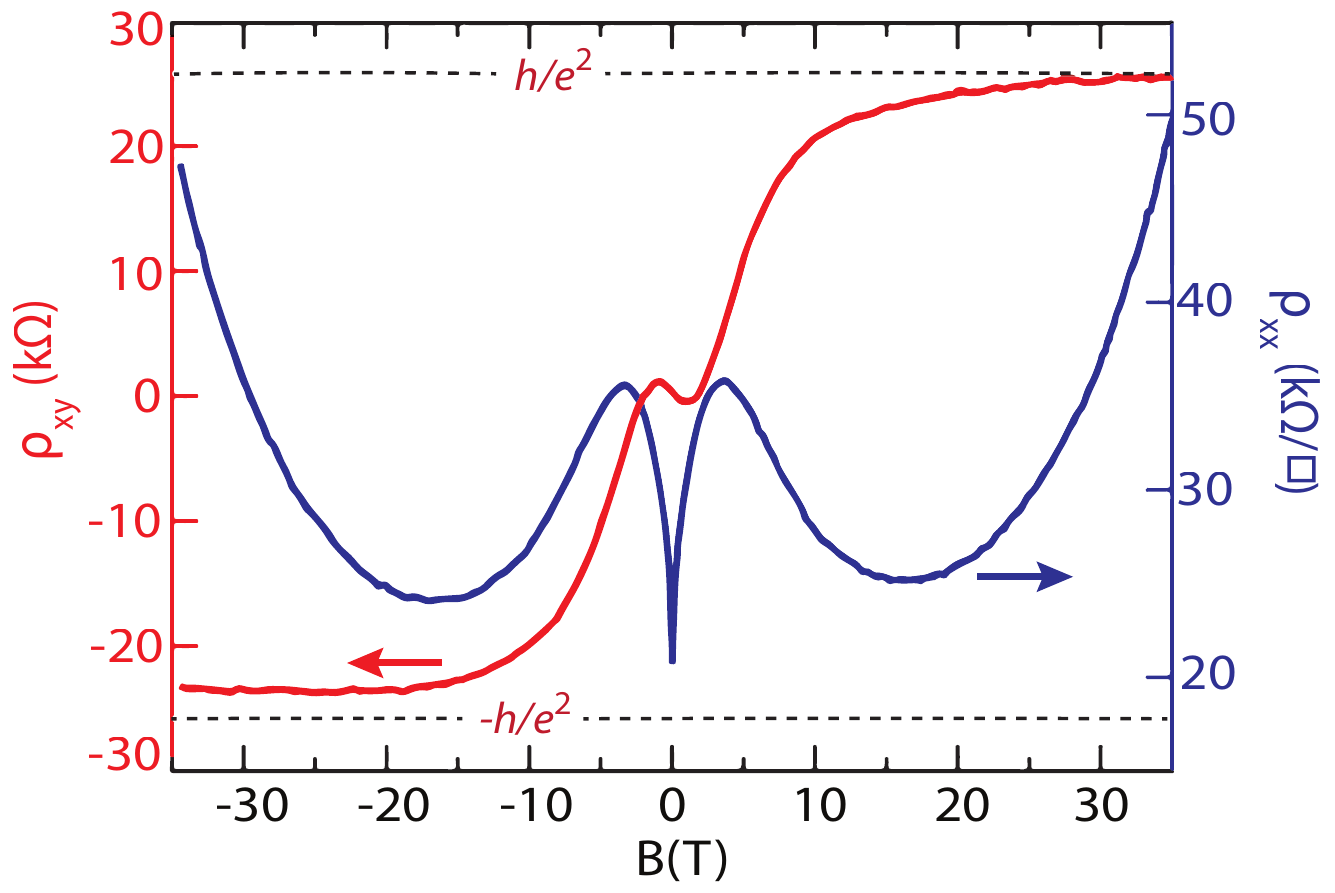}
\caption{\label{fig:R6mix} QHE in a mixed carrier type sample. The Hall (left axis) and longitudinal (right axis) resistivities of a (gate-voltage) device in a material having a high-mobility n-type carrier and a lower mobility p-type carrier, so that the Hall slope is negative at $\left|B\right|$ $<$ 0.99 T and positive at higher fields ($V_g$ = -20 V). The material becomes Landau quantized at higher fields, reaching a p-type $\nu$=1 plateau at around B = 20 T.}
\end{figure}
\newpage
\begin{figure}[H]
\centering
\includegraphics[scale=0.69]{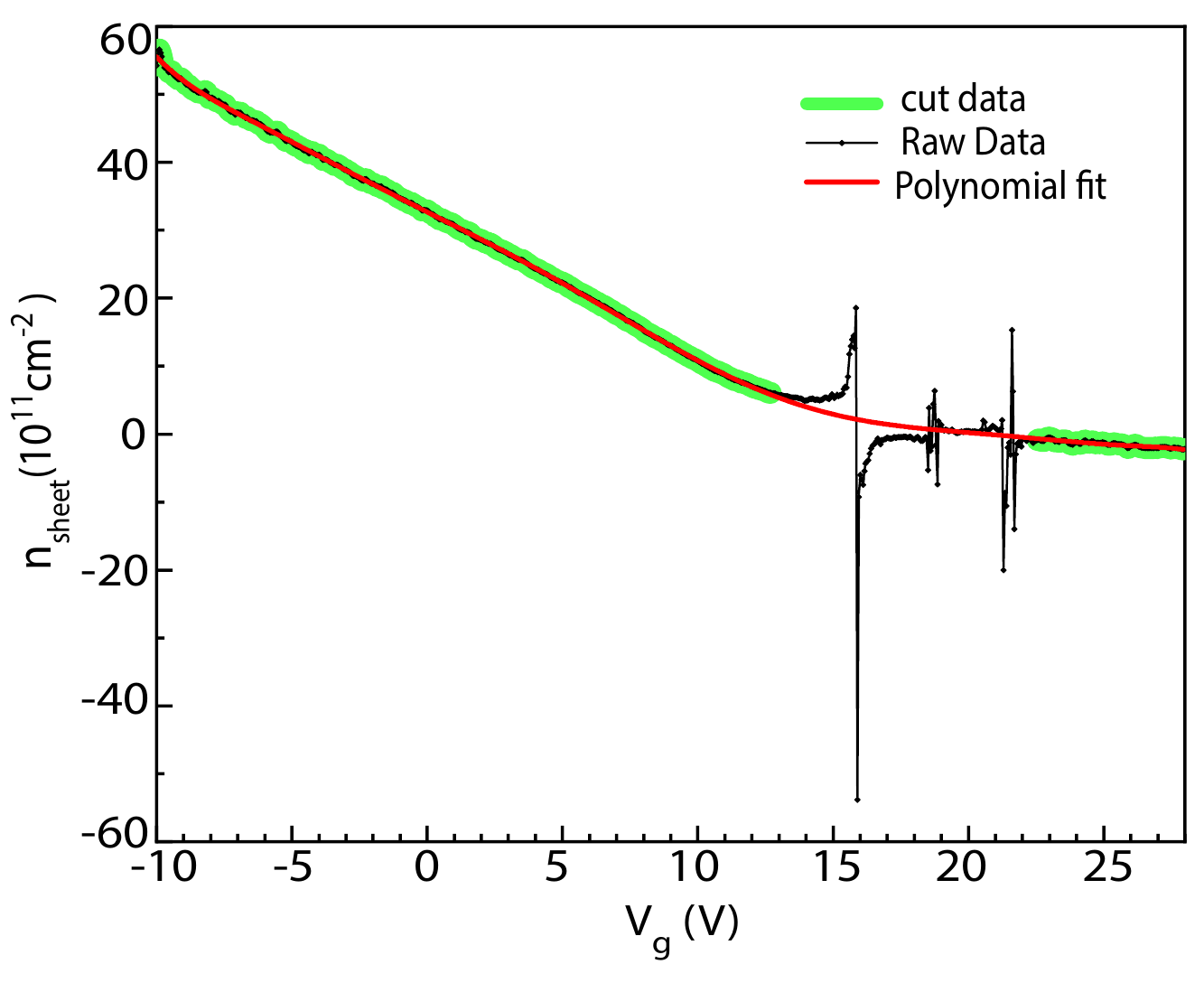}
\caption{\label{fig:nv} The black line is the raw data for \textit{n$_{sheet}$} as a function of gate-voltage, where the top gate of the gated device was swept from -10 V to 28 V at fields between B = -1 T to 1 T. At each gate voltage, the Hall resistance was found by linear fitting, and was used to find the carrier density. The jumps in the raw data is due to mixed carrier density near CNP which artificially gives a flat Hall slope. The Hall resistance fit quality near charge neutrality, however, is poor due to mixing between the Hall and longitudinal resistivities. Therefore, to get a sensible relationship between the carrier density and gate voltage we exclude the bad part of the data and use the remaining data (green curve) with a polynomial fit. The red curve is a 9$^{th}$ degree polynomial fit to the data.}
\end{figure}
\newpage
\begin{figure}[H]
\centering
\includegraphics[scale=0.51]{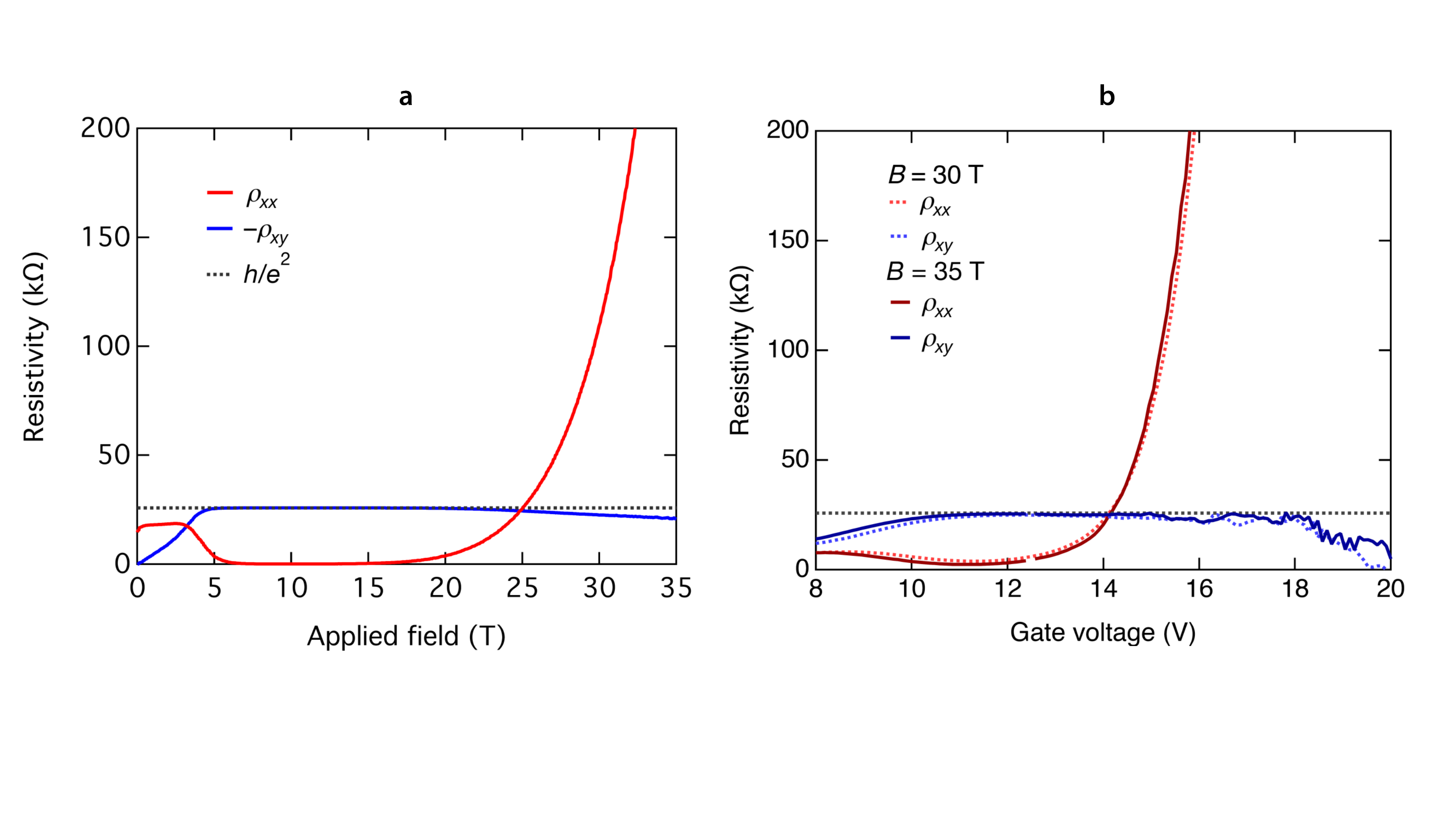}
\caption{\label{fig:Hallinsul}a) The longitudinal (red) and Hall (blue) resistivities of N8.1.5\%; at high fields at 300 mK, shown after (anti-)symmetrization between positive and negative signed fields. A dashed line indicates the value of the resistance quantum $h/e^2$. A $\nu$ = 1 quantum Hall plateau is observed at applied fields between approximately 5 and 20 T.  The quantum Hall to insulator transition occurs at $B_c$ = 24 T. For clarity, the sign of the Hall resistivity has been reversed in this figure. b) The longitudinal (red) and Hall (blue) resistivities (after (anti-)symmeterization) of the gated device at 30 T (dashed lines) and 35 T (solid lines) as a function of gate voltage at 300 mK. A $\nu$ = 1 quantum Hall plateau is observed centered around $V_g$ = 11.5 V (at 30 T) and $V_g$ = 11.3 V (at 35 T). The $\rho_{xx}$ diverges at higher gate voltages as carriers are depleted, while the $\rho_{xy}$ remains roughly quantized at $h/e^2$  up to around $V_g$ = 18 V, at which  $\rho_{xx}$ = 1.2 M$\ohm$. This observation suggests the system may be a quantized Hall insulator in approximately the range $V_g$ = 14 V to 18 V.}
\end{figure}

\newpage

\begin{table}[H]
\centering
\caption{Ti vapor pressure as a function of temperature provided by Veeco or on line. }
{\small
\begin{tabular}{ |c|c|c|c|c|c|c|c| } 
 \hline
Pressure (Torr) & 10$^{-8}$ & 10$^{-7}$  & 10$^{-6}$  & 10$^{-5}$  & 10$^{-4}$  & 10$^{-3}$  \\ 
\hline
Temperature ($\degree$C) & 1062 & 1137 & 1227 & 1327 & 1442 & 1577 \\ 
 \hline
\end{tabular}
}
\end{table}

\vspace{1cm}	 

\begin{table}[H]
\centering
\caption{Some of the Ti fluxes at lower temperatures along with corresponding vapor pressure and calculated doping level which were used for the samples in this work. For calculating Ti flux at different temperatures, we first start by fitting the vapor pressure data as a function of temperature, with a polynomial  $\ln(P)=a_1+a_2T+a_3 T^2$. The fitting parameters $a_1= -59.6, a_2= 0.051$, and $a_3= -1.11 \times 10^{-5}$ can thus be extracted.  Also for an ideal gas, it can be shown that $\Phi_i=\Phi_j\cdot \frac{P_i}{P_j}\left(\frac{T_j}{T_i}\right)^{1/2}$= $\Phi_j\cdot \frac{a_1+a_2T+a_3 T^2}{a_1+a_2T+a_3 T^2} \left(\frac{T_j}{T_i}\right)^{1/2}$
where  $\Phi_i$ and $\Phi_j$ are source fluxes (atoms/$cm^{2} \cdot s$), $P_i$ and $P_j$  are vapor pressures of the source (Torr) for the source temperatures, $T_i$ and $T_j$ (K), respectively. Therefore, by knowing the flux $\Phi_j$  at a specific temperature (usually it is measured at higher temperatures where you have reasonably measurable flux by the quartz crystal microbalance (QCM) system), $\Phi_i$ can be calculated. Here, for each doped sample, we systematically vary Ti flux (temperature) while keeping Sb at a constant flux/temperature. The doping level is based on  $\frac{\Phi_{\text{Ti}}}{\Phi_{\text{Ti}}+\Phi_{\text{Sb}}}\times 100\%$. $\Phi_{\text{Sb}}$ $\approx$ $1.2\times 10^{13} cm^{-2}s^{-1}$ at 395 $\degree$C and 
 $\Phi_{\text{Ti}}$ $\approx$ $6\times 10^{12} cm^{-2} s^{-1}$ at 1550 $\degree$C.
}
\begin{tabular}{ |c|c|c|c| } 
 \hline
Ti temperature ($\degree$C) & Vapor pressure ($\mu$Torr)  & Flux (10$^{10}$cm$^{-2}$S$^{-1}$)  & Doping level (\%) \\ 
\hline
1325 & 9.26 & 8.95 & 0.7 \\ 
\hline
1340 & 12.7 & 12.3 & 1.0 \\ 
\hline
1344 & 13.9 & 13.3 & 1.1 \\
\hline
1350 & 15.7 & 15.1 & 1.3 \\
\hline
1360 & 19.4 & 18.5 & 1.5\\
\hline
\end{tabular}
\end{table}

\newpage


\section{Supplementary Text (Theory)}

In this section, we start by briefly discussing the low energy effective model of topological insulators and Dirac surface states and their Landau level spectrum in the presence of a strong magnetic field. Next, we explain the effect of random impurity potential on the energy spectrum and transport measurements.

\subsection{Bulk effective model of topological insulators}
We use the  four-band Dirac Hamiltonian to study the low energy properties of Sb$_2$Te$_3$ thin films. This model was introduced in Refs.\cite{ZhangBiSe_2009,Liu_4band}. 
Near the $\Gamma$ point, the effective low energy Hamiltonian is written as
\begin{align} \label{eq:Dirac}
H({\bf k})=\epsilon({\bf k})\mathbb{I}+ v_F( k_x\Gamma_1+  k_y\Gamma_2) + v_3 k_z\Gamma_3+ {\mathcal M}({\bf k}) \Gamma_4
\end{align}
where the Dirac matrices are given by
\begin{align*}
&\Gamma_s=  \tau_1\otimes \sigma_s, 
\qquad \Gamma_4=    \tau_3\otimes\mathbb{I}_2,
\end{align*}
and $s=1, 2, 3$. In this convention the $\sigma_{1,2,3}$ and $\tau_{1,2,3}$ matrices act on the spin and orbital degrees of freedom respectively. Moreover, $\epsilon({\bf k})=C_0+C_1 k_z^2+C_2 k_\perp^2$ and ${\mathcal M}(k)=M_0+M_1 k_z^2 +M_2 k_\perp^2$. 
The parameters $v_F$, $v_3$, $C_i$ and $M_i$ can be chosen carefully to reproduce the band structure near the $\Gamma$ point of the Sb$_2$Te$_3$~\cite{Liu_4band}. 
We should note that  $M_1$ and $M_2$ coefficients are positive and  $M_0$ is negative 
in the Bi$_2$Se$_3$ family of materials, e.g.~Bi$_2$Se$_3$, Bi$_2$Te$_3$ and Sb$_2$Te$_3$.

In the presence of a magnetic field, the canonical momenta is modified into
\begin{align}
D_j= p_j - eA_j,
\end{align}
which satisfy the commutation relation, $[D_x,D_y] = i {\hbar^2}/{\ell_B^2}$ where $\ell_B^2=\hbar/eB$.
Let us introduce the ladder operators
\begin{align}
a=\frac{\ell_B}{\sqrt{2}\hbar}(D_x + i D_y ), \qquad a^\dag=\frac{\ell_B}{\sqrt{2}\hbar}(D_x - i D_y ),
\end{align}
which obey $[a,a^\dag]=1$. Then, the TI Hamiltonian in the presence of magnetic field becomes
\begin{align}\label{eq:H_LL}
H(k_z)= \widetilde{\epsilon}\ \mathbb{I}_4 +\left(
\begin{array}{cccc}
\widetilde{\mathcal M}& v_3 k_z & 0 & v_F  \frac{\sqrt{2}}{\ell_B} a \\
 v_3 k_z & -\widetilde{\mathcal M} & v_F \frac{\sqrt{2}}{\ell_B} a & 0 \\
  0 & v_F \frac{\sqrt{2}}{\ell_B} a^\dag & \widetilde{\mathcal M} & -v_3 k_z \\
 v_F \frac{\sqrt{2}}{\ell_B} a^\dag & 0 & -v_3 k_z & -\widetilde{\mathcal M}
\end{array}
\right),
\end{align}
where
$\widetilde{\mathcal M}(k_z,a^\dag a)=M_0+M_1k_z^2 + 2 M_2/\ell_B^2 (a^\dag a+ 1/2)$ and
$\widetilde{\epsilon}(k_z,a^\dag a)=C_0+C_1k_z^2 + 2 C_2/\ell_B^2 (a^\dag a+ 1/2)$. 
Next, we bring the Hamiltonian into block diagonal form using the basis
\begin{align}
\ket{\Psi_N}= (\alpha_{N-1} \ket{N-1},\beta_{N-1} \ket{N-1},\alpha_{N} \ket{N},\beta_{N} \ket{N})^T,
\end{align}
that is
\begin{align} \label{eq:bulk_LL}
H(k_z,N)=   \left(
\begin{array}{cccc}
\widetilde{\mathcal M}_{N-1}^+  & v_3 k_z & 0 &  v_F\frac{\sqrt{2N}}{\ell_B}  \\
 v_3 k_z &\widetilde{\mathcal M}_{N-1}^- &   v_F\frac{\sqrt{2N}}{\ell_B}  & 0 \\
  0 &   v_F\frac{\sqrt{2N}}{\ell_B}  &\widetilde{\mathcal M}_{N}^+  & -v_3 k_z \\
 v_F\frac{\sqrt{2N}}{\ell_B}  & 0 & -v_3 k_z & \widetilde{\mathcal M}_{N}^-
\end{array}
\right)
\end{align}
where $\ket{N}$ is the eigenstate of the Harmonic operator $a^\dag a \ket{N}=N\ket{N}$
and $\widetilde{\mathcal M}_{N}^\pm= \widetilde{\epsilon}(k_z,N)\pm\widetilde{\mathcal M} (k_z,N)$. The zeroth Landau levels must be considered separately. They are given in the basis $\ket{\Psi_0}=(0, 0,  \alpha_0 \ket{0}, \beta_0\ket{0})^T$.

\subsection{Surface effective model}

In this part, we discuss the surface Dirac Hamiltonian ~\cite{Konig2008,Zhou_surface_2008,Linder2009,Lu_surface_2010,Shan_surface_2010,Liu_4band} and derive the Landau spectrum in the presence of magnetic field.

First, we need to find the zero modes which form the Dirac node on the surfaces.
We model the top (bottom) surfaces by imposing an open boundary condition along the $z$-direction. This boundary condition breaks the translational symmetry and hence, $k_z$ is no longer a good quantum number and must be replaced by the gradient operator $-i\partial_z$  in  real space. 
So, the zero modes are the solutions to the following Schr\"oedinger equation,
\begin{align}
\left[-i \sigma_3 \tau_1   v_3  {\partial_z}+ \tau_3 (M_0-M_1\partial_z^2) \right]|\Psi\rangle =0.
\end{align}
Since the above operator is diagonal in the spin basis, solutions can be written in the up/down spin states. Hence, we need to solve the two component equation
\begin{align} \label{eq:zeromode_B0}
\left(
\begin{array}{cc}
M_0 - M_1 \partial_z^2 & \mp i v_3 \partial_z \\
\mp i v_3  \partial_z & -(M_0 - M_1 \partial_z^2)
\end{array}
\right) |\psi_p\rangle=0
\end{align}
In principle, the above equation can be solved for a system of thickness $d$ and the finite-size tunneling gap in the spectrum can be computed~\cite{Konig2008,Zhou_surface_2008,Linder2009}.
Then, we do perturbation theory to the in-plane kinetic terms in the basis of zero modes and derive the surface Hamiltonian (normal to $z$ direction). We shall only quote the result. The surface Hamiltonian is found to be
\begin{align} \label{eq:surf_ham}
H_{\text{surf}}=v_F \eta_3 \otimes (k_x \sigma_2 - k_y \sigma_1) + \Delta_t (k_\perp)\ \eta_1 \otimes \mathbb{I}_2,
\end{align}
where $\eta_i$ are Pauli matrices in top (bottom) surface, $ \Delta_t (k_\perp) =\Delta_0 + \Delta_2 k_\perp^2$ denotes a tunneling amplitude between top and bottom surfaces ($k_\perp^2=k_x^2+k_y^2$). For a thick sample, the surface states decay exponentially into the bulk, where the characteristic length is  
\begin{align} \label{eq:penetration}
\xi_{\pm}^{-1} = \text{Re}\left[ \frac{v_3 \pm \sqrt{v_3^2+ 4 M_0 M_1} }{2M_1} \right].
\end{align} 
The lowest order term, which gives the inter-plane tunneling, comes from the kinetic term as in
\begin{align} \label{eq:tunneling}
\Delta_0 \approx M_1 \frac{L}{\xi^3} \  e^{-L/\xi}, \qquad
\Delta_2\approx M_2 \frac{L}{\xi}  \  e^{-L/\xi}.
\end{align}
Using the parameters of Liu~\emph{et.~al.}~\cite{Liu_4band},
$M_0\approx -0.22\ $eV, $M_1\approx 20\ $eV\AA$^2$, $M_2\approx 50\ $eV\AA$^2$,  
$v_3\approx 0.84\ $eV\AA, and $v_F\approx 3.40\ $eV\AA
, we obtain $\Delta_0\approx 6 m$eV and $\Delta_2\approx 15$eV\AA$^2$ for a sample of thickness $L=8nm$. We note that this calculation overestimates the finite-size energy gap $\Delta_0$, which is experimentally measured to be $\Delta_0^{\text{exp}}\approx 0.15m$eV.

In the presence of strong magnetic field, the zero energy modes on the surface are found by using the bulk zeroth Landau level Hamiltonian (\ref{eq:bulk_LL}),
\begin{align}  \label{eq:zeromode_B}
 H(-i\partial_z, N=0) = 
\left(
\begin{array}{cc}
\widetilde{\cal M} & i v_3 \partial_z  \\
i v_3  \partial_z & -\widetilde{\cal M}
\end{array}
\right)
\end{align}
in which $\widetilde{\cal M}= M_0+M_2/\ell_B^2 -M_1 \partial_z^2$.
Note that compared with the zero-field Hamiltonian (\ref{eq:zeromode_B0}), the additional term $M_2/\ell_B^2$ makes the effective bulk gap smaller, and hence increases the penetration depth $\xi$ in (\ref{eq:penetration}). 
Given the parameter values mentioned earlier, we get $M_2/\ell_B^2\lesssim 16$meV for $B=20 T$. This in turn leads to a negligible increase in the tunneling gap $\Delta_0$ of a sample with $L=8nm$ thickness.
 
Furthermore, the quadratic term $\Delta_2 (k_x^2+k_y^2)$ in the presence of magnetic field becomes 
 $2\Delta_2/\ell_B^2 (N+1/2)$. Hence, for the zeroth Landau level $N=0$, it increases the tunneling gap $\Delta_0$ by $\Delta_2/\ell_B^2 \lesssim 5 m$eV. This effect is much smaller than the estimation of Ref.~\cite{Shen_LLshift}, since $\Delta_2$ in our case is much smaller. Finally, the upper bound to the change in $\Delta_0$ is $5 m$eV.


 \subsubsection{Surface Landau levels}

Here, we study the Landau spectrum of the surface states. We consider the generic Hamiltonian (with already modified parameters),
\begin{align} \label{eq:surf_ham_LL}
H_{\text{surf}} &= v_F \eta_3 \otimes (D_x \sigma_2 - D_y \sigma_1) + {\Delta_0} \ \eta_1\otimes \mathbb{I}_2 + \Delta_z  \mathbb{I}_2 \otimes \sigma_3 \nonumber \\
& = \left(
\begin{array}{cccc}
\Delta_z &  -i \omega_0  a^\dag & {\Delta_0} & 0 \\
i \omega_0  a  & -\Delta_z  & 0 & {\Delta_0} \\
 {\Delta_0} & 0 & \Delta_z & i \omega_0  a^\dag  \\
0 & {\Delta_0} & -i\omega_0 a   &- \Delta_z
\end{array}
\right)
\end{align}
where $\omega_0=v_F \sqrt{2}/\ell_B$,
is the energy scale of Landau levels and we also added a Zeeman term $\Delta_z$.
The near zero Landau levels are given by
\begin{subequations} \label{eq:surf_LL_evals}
\begin{eqnarray} 
\ket{0,+}&=\frac{1}{\sqrt{2}} (\ket{t_0}+\ket{b_0}), \qquad \varepsilon_+= \Delta_z + \Delta_{0} \\
\ket{0,-}&=\frac{1}{\sqrt{2}} (\ket{t_0}-\ket{b_0}), \qquad \varepsilon_-= \Delta_z - \Delta_{0}
\end{eqnarray}
\end{subequations}
where 
\begin{align}
\ket{t_{0}} &=(\ket{0}, 0, 0, 0)^T, \\
\ket{b_{0}} &=(0,0,\ket{0}, 0)^T,
\end{align}
are zeroth Landau levels on top and bottom surfaces, respectively. The eigenstates are simply (anti-)bonding combinations of the surface zeroth Landau levels. It is important to note that the Zeeman field shifts both lowest Landau levels (LLLs) in the same direction and does not change the tunneling gap between LLLs.

For higher landau levels, we use the basis
\begin{align}
\ket{t_N^+}&=(\ket{N},i\ket{N-1},0,0)^T,   \\
\ket{t_N^-}&=(\ket{N},-i\ket{N-1},0,0)^T,    \\
\ket{b_N^+} &=(0,0,\ket{N},-i\ket{N-1})^T  \\
\ket{b_N^-} &=(0,0,\ket{N},i\ket{N-1})^T  
\end{align}
where the Hamiltonian in this subspace reads
\begin{align}
H_N= \omega_0 \sqrt{N} \mathbb{I}\otimes \alpha_3+ {\Delta_0} \eta_1\otimes \alpha_1 + \Delta_z \mathbb{I} \otimes \alpha_1,
\end{align}
and $\alpha_i$'s are a set of Pauli matrices.
The energy spectrum is given by
\begin{align} \label{eq:landaulevels}
\pm \varepsilon_{N}= \pm \sqrt{N \omega_0^2 + (\Delta_z \pm \Delta_0)^2 }.
\end{align}
It is evident from the above expression that the energy shifts associated with the Zeeman field and tunneling are smaller for higher LLs with larger $N$.

\subsection{Effect of disorder on surface states}

In this part, we study the effect of random impurity potential on surface states both in the presence and absence of a magnetic field. The competition between the magnetic field and random disorder potential can be characterized in terms of a dimensionless parameter $\Gamma/\hbar\omega_0$ which roughly speaking, compares the LL broadening and the LL spacing. 
Our results are summarized as $2$d phase diagrams in Figs.~\ref{fig:phasediag_m}{b} and~\ref{fig:phasediag}{b} where the corresponding zero-field systems are described by two massless Dirac cones and massive Dirac Hamiltonians  due to the inter-surface tunneling gap, respectively.
The important observation in either case is that the $\nu=1$ plateau is quite robust even in the strong disorder limit.

We model impurities (crystal defects, charged defects, etc.) by adding a random potential to the clean Hamiltonian of surface states (\ref{eq:surf_ham}),
\begin{align}
H= H_{\text{surf}} + V_{\text{rand}}
\end{align}
where  
\begin{align}
V_{\text{rand}}= \int d\textbf{r}\ v(\rv)\ \psi^\dag(\rv)  \psi(\rv),
\end{align}
 and $v(\rv)$  is the impurity potential profile which is given as a set of uncorrelated random numbers.

\subsubsection{Landau-level broadening}
\begin{figure}
\includegraphics[scale=0.28]{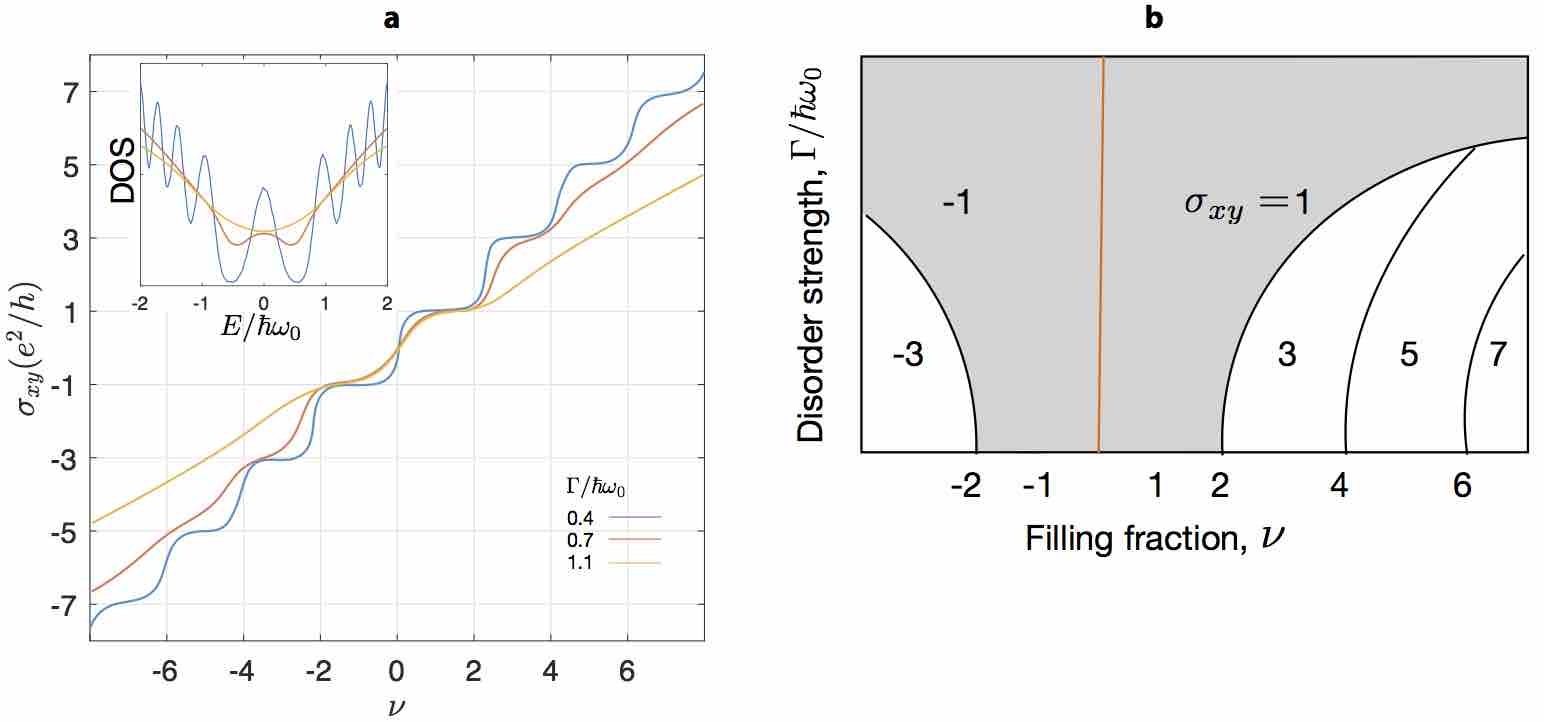}
\caption{\label{fig:phasediag_m}a) Hall conductance (Inset: Density of states) for various disorder strengths. b) Phase diagram of disordered two decoupled Dirac Landau levels ($\Delta_0/\omega_0=0$). 
Other parameters are  $N_{\text{imp}}/L^2=1/2$ and $d/\ell_B=0.7$. Each area is denoted by its corresponding value of the Hall conductance $\sigma_{xy}$.
The orange line indicates the critical line between the $\nu=1$ and $\nu=-1$ plateaus.
This result was reproduced based on the Ref.\cite{Nomura2008}.}
\end{figure}

Discrete Landau levels (\ref{eq:landaulevels}) are broadened due to scattering caused by the disorder potential. Following~\cite{Ando1984,Nomura2008}, we use the long range disorder potential profile
\begin{align} \label{eq:Vdis}
v(\textbf{r})= \sum_{j=1}^{N_{\text{imp}}} \frac{u_j}{2\pi d^2} \exp(-|\textbf{r}-\textbf{R}_j|^2/2d^2),
\end{align}
which consists of $N_{\text{imp}}$ impurities at random locations ${\bf R}_j$. To ensure the neutrality, we assume equal number of positive and negative potential energies $u_j=\pm u$.
A measure of disorder strength is defined in terms of the LL broadening parameter,
\begin{align}
\Gamma^2=8\pi u^2 \frac{N_{\text{imp}}}{(\ell_B^2+2d^2)L^2},
\end{align}
derived from self-consistent Born approximation~\cite{Ando1984}. The quantity $\Gamma/\hbar\omega_0$ is then gives a relative ratio between LL spacing and the bandwidth. As the inset of Fig. \ref{fig:phasediag_m}a shows, when $\Gamma/\hbar\omega_0$ is small (e.g., the blue curve $\Gamma/\hbar\omega_0=0.4$), the DOS preserves its discrete form and we see the LLs are well separated. This corresponds to the presence of Hall plateaus at quantized values for  the Hall conductance (e.g., blue curve in Fig. \ref{fig:phasediag_m}a). As we go to larger $\Gamma$, the LL mixing increases and in the extreme limit, the DOS becomes quite smooth (as in the yellow curve for  $\Gamma/\hbar\omega_0=1.1$) and the Hall plateaus except for the $\nu=1$ are destroyed.

\subsubsection{Finite-size tunneling}

The top and bottom surfaces of a thin-film TI are always coupled via tunneling through the bulk (the tunneling amplitude however could be exponentially small, see Eq.(\ref{eq:tunneling})).
This results in a small gap in the Dirac surface spectra. In principle, we expect to observe an insulating behavior near the charge neutrality point within the surface energy gap. However,  as we see in this part random disorder smears the tunneling gap and drives the gapped system towards a metal. We investigate the gap smearing phenomena in the presence of strong magnetic field and
our observation is summarized as follows: In the weak magnetic field limit (or strong disorder) $\Gamma/\hbar\omega_0\gg1$
 the finite-size tunneling gap is smeared into a critical metallic region, while in the strong magnetic field (or weak disorder) $\Gamma/\hbar\omega_0\ll1$, we get an insulating phase between the two zeroths LLs which are separated by the finite-size tunneling gap.

We start by studying the zero-field limit analytically.
We assume that $V(\rv)$ are random numbers taken from a uniform (or Gaussian) distribution $[-W,W]$ and satisfy
\begin{align}
\overline{V(\rv)}=0, \qquad  \overline{V(\rv) V(\rv')}= \frac{W^2}{12} \delta(\rv-\rv'),
\end{align}
where overbar denotes the disorder average

Our goal is to see the band smearing in the presence of disorder.
Because of disorder, the ensemble averaged propagator develops a self-energy part $\Sigma(i\omega_n)$ as in 
\begin{align}
{\cal G}(\kv,i\omega_n)=  \frac{1}{i\omega_n - h(\textbf{k})-\Sigma(i\omega_n)}.
\end{align}
It is this term $\Sigma(i\omega_n)$ which could renormalize the original gap in $h(\textbf{k})$ and if it has an imaginary time it gives rise to scattering time
\begin{align}
\tau=\lim_{\omega_n\to 0} 1/\text{Im}[\Sigma(\omega_n)].
\end{align}
The scattering time $\tau$ naturally leads to a mean-free path $l= v_F \tau$.
In the limit $\tau <\infty$, the conduction becomes diffusive, otherwise the conduction is pseudo-ballistic.  
To approximate the self-energy, we use the self-consistent Born approximation
\begin{align}
\Sigma(i\omega_n)=& \frac{W^2}{12L^d} \int \frac{d\textbf{k}}{i\omega_n - h(\textbf{k})-\Sigma(i\omega_n)},
\end{align}
which is equivalent to the propagator in the replica limit (non-crossing diagrams).

Let us plug in the surface Hamiltonian (\ref{eq:surf_ham}), and find the self-energy.
The self-energy can be decomposed as
$\Sigma = \Sigma_0 \mathbb{I} + \Sigma_1 \eta_1$,
where $\Sigma_0=\Sigma_{0r} + i \Sigma_{0i}$ is complex-valued. The self-consistent equation for $\Sigma_1$ is given by
\begin{align}
\Sigma_1= - (\Delta_0+ \Sigma_1) \frac{\pi W^2}{12 L^2} \log\left(\frac{(v_F\Lambda)^2}{(\Delta_0+ \Sigma_1)^2}+1\right),
\end{align}
in which we neglected $\Delta_2$ term. We also introduce the 3d bulk gap $\Lambda\sim M_0$ as a high energy cut-off. The crucial message is that $\Sigma_1<0$, which implies that the renormalized hybridization gap $\Delta_0+ \Sigma_1$ is smaller than the original gap $\Delta_0$. Physically, this is due to the fact that the bands are being smeared by disorder~\cite{Mong2012} and the system is driven towards a gapless metallic phase. 
In the above discussion, we use short range delta correlated disorder to derive this result analytically. However, it also holds for other types of scalar disorder potential.

\begin{figure}
\includegraphics[scale=0.51]{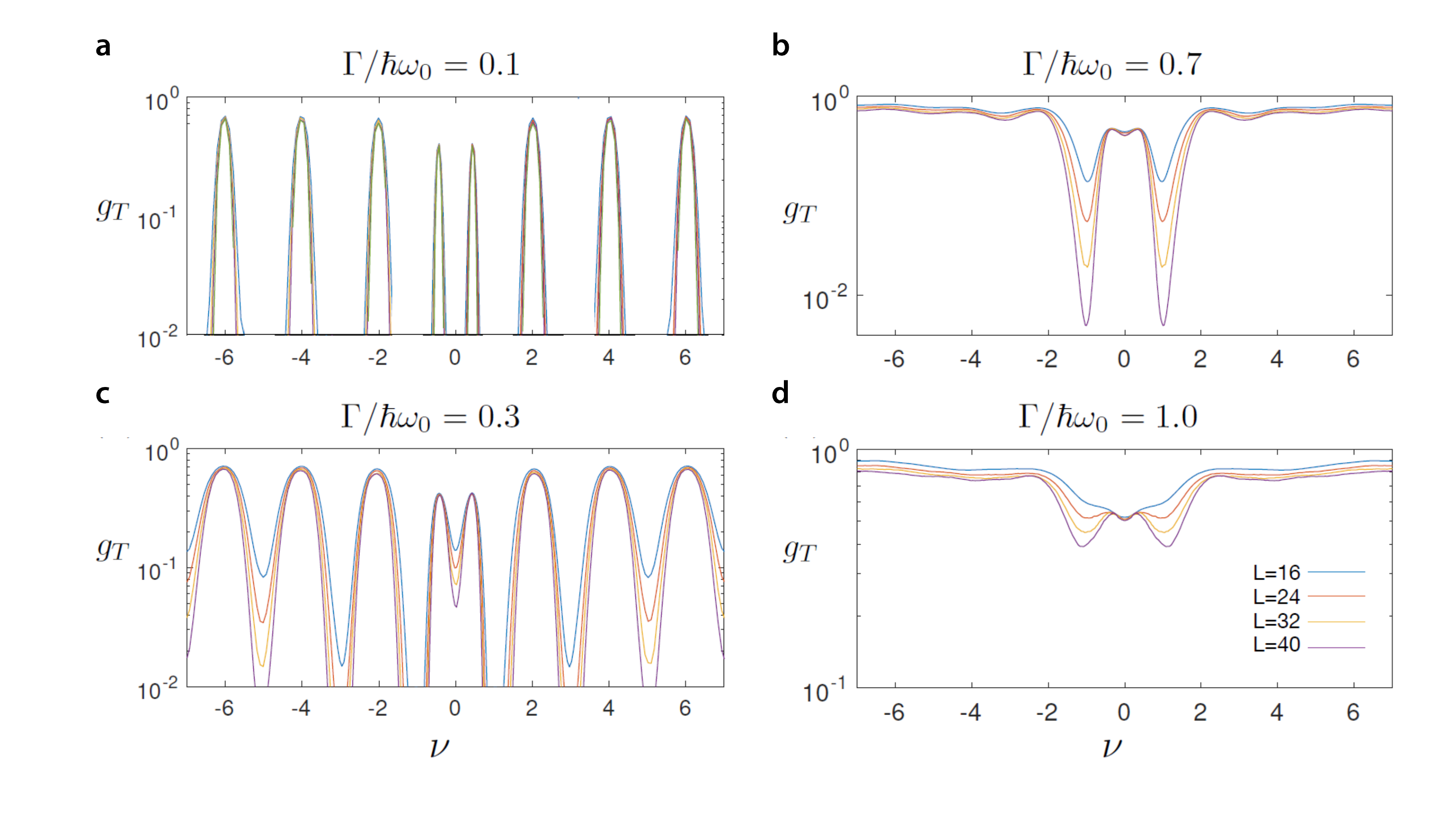}
\caption{\label{fig:sxx}  Thouless number as a measure of longitudinal conductivity for various disorder strength $\Gamma$ a) 0.1, b) 0.3, c) 0.7,and d) 1.0. The parameters are  $\Delta_0/\omega_0=0.1$, $N_{\text{imp}}/L^2=1/2$, and $d/\ell_B=0.7$. }
\end{figure}

\begin{figure}
\includegraphics[scale=0.08]{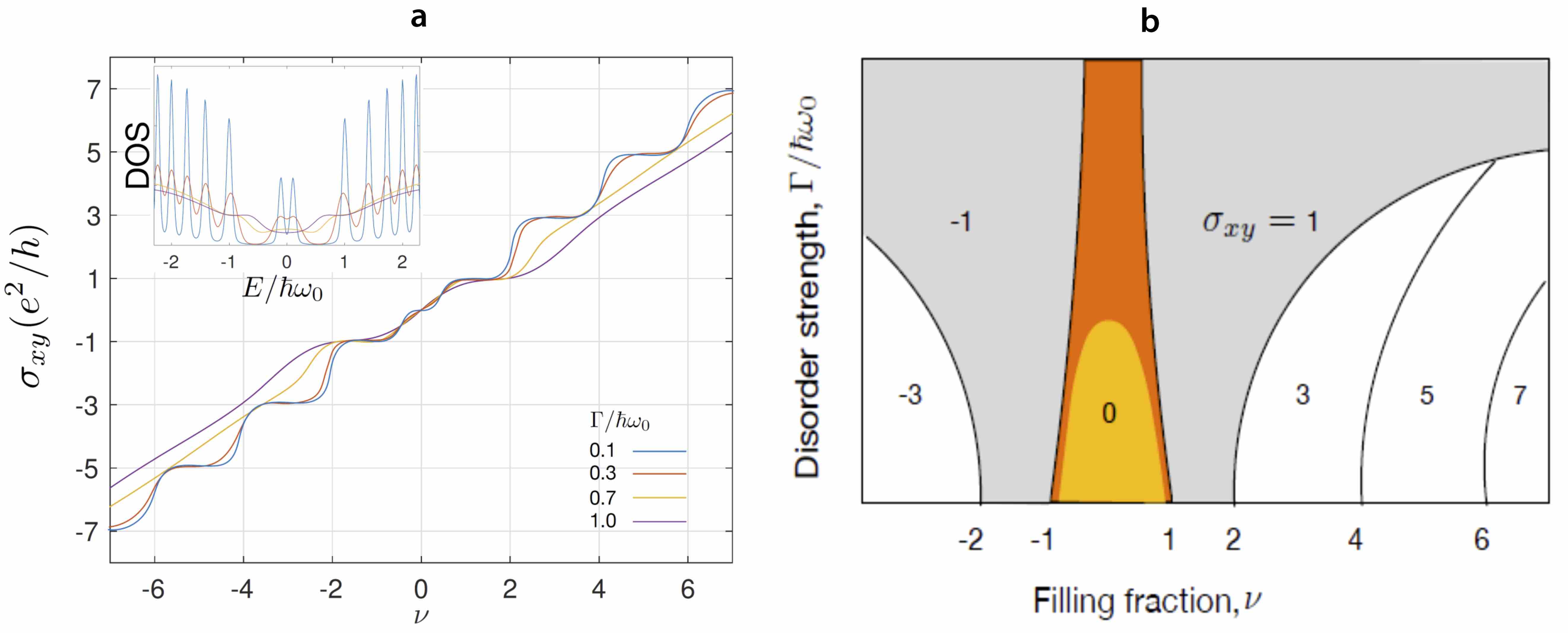}
\caption{\label{fig:phasediag}a) Hall conductance (Inset: Density of states.) for various disorder strengths. b) Phase diagram of disordered two tunnel-coupled Dirac Landau levels. Each area is denoted by its corresponding value of the Hall conductance $\sigma_{xy}$.
In particular, the dark and light orange regions indicate the critical metallic region (driven by disorder) and the Hall insulator (formed in the strong magnetic field limit), respectively.
The parameters are the same as in Fig. \ref{fig:sxx}. }
\end{figure}

Next, we study the effect of disorder on LLs associated with gapped Dirac surface states. Analytical calculations in this case are rather tedious and we resort to numerical investigations. We first compute the Thouless number to determine the fate of the insulating phase near the charge neutrality point. The Thouless number is a measure of longitudinal conductivity and is defined by
\begin{align}
g_T= \frac{\braket{\Delta \varepsilon}}{\delta \varepsilon},
\end{align}
where $\Delta \varepsilon$ is  the energy shift induced by changing the boundary condition from periodic to anti-periodic. The mean level spacing is $\delta \varepsilon=1/(L^2 \rho(\varepsilon))$ in terms of DOS $\rho(\varepsilon)$. The average energy shift is evaluated by $\braket{\Delta \varepsilon}=\exp(\overline{\ln\Delta \varepsilon})$. 
Figure~\ref{fig:sxx}a shows the results for various disorder strengths. In each panel, $g_T$ is plotted as a function of filling fraction $\nu$ for different system sizes $L \times L$.
The peaks of $g_T$ indicate the center of LLs where the extended (delocalized) states reside. 
In fact, these peaks represent the critical metal (i.e., transition point) in the plateau-to-plateau transitions where the longitudinal conductivity is scale invariant (i.e.,  $g_T$ does not change with system size).
We should note that the peaks are located at even filling fractions $\nu=2n$ in Fig. \ref{fig:sxx}. 
The reason is here we set $\Delta_z=0$ as our focus is mostly the LLL physics which are just shifted by the Zeeman term (c.f.~Eq.(\ref{eq:surf_LL_evals})). $\Delta_z=0$ implies that the higher LLs are two-fold degenerate as seen in Eq.(\ref{eq:landaulevels}). 
Between the peaks in Fig. \ref{fig:sxx}, we observe that $g_T$ decreases as we increase the system size. This is a hallmark of an insulating behavior. In other words, $g_T$ curves in each panel of Fig. \ref{fig:sxx}a shows a series of insulating regions separated by critical metals. These insulating regions in Fig. \ref{fig:sxx} are the integer quantum Hall (IQH) plateaus as shown in Fig.  \ref{fig:phasediag}a.

 Let us look more closely at the evolutions of $g_T$ peaks in the four panels of Fig. \ref{fig:sxx}. For weak disorder ($\Gamma/\hbar\omega_0=0.1$ and $0.3$), the peaks are sharper and there are two peaks which separate the $\nu=0$ plateau from $\nu=\pm 1$ plateaus (see also blue curve of Fig. \ref{fig:phasediag}a). As the disorder strength is increased ($\Gamma/\hbar\omega_0=0.7$ and $1.0$), the peaks become wider. The insulating IQH plateaus between the peaks at higher filling fractions start to disappear while the $\nu=\pm 1$ plateaus still persist (see violet curve of Fig. \ref{fig:phasediag}a). This is an evidence for the robustness of $\sigma_{xy}=1$ plateaus. Moreover, the $\nu=0$ insulating phase seem to show a scale-invariant behavior at larger disorder strength. This  means that the original $\nu=0$ plateau has turned into a critical metal. 
 
 We summarize our results in terms of a phase diagram in Fig. \ref{fig:phasediag}b.  The destruction of  higher LL plateaus which we also observe in our numerics is the well-known levitation phenomena where  the extended states are effectively pushed to higher energies by disorder~\cite{Khmelnitskii,Laughlin}. We believe that the nature of the transition from $\nu=0$ plateau into a critical metal is similar to the disorder smearing effect explained for the zero-field regime.  
 
 \newpage
 
\begin{figure}[H]
\centering
\includegraphics[scale=0.53]{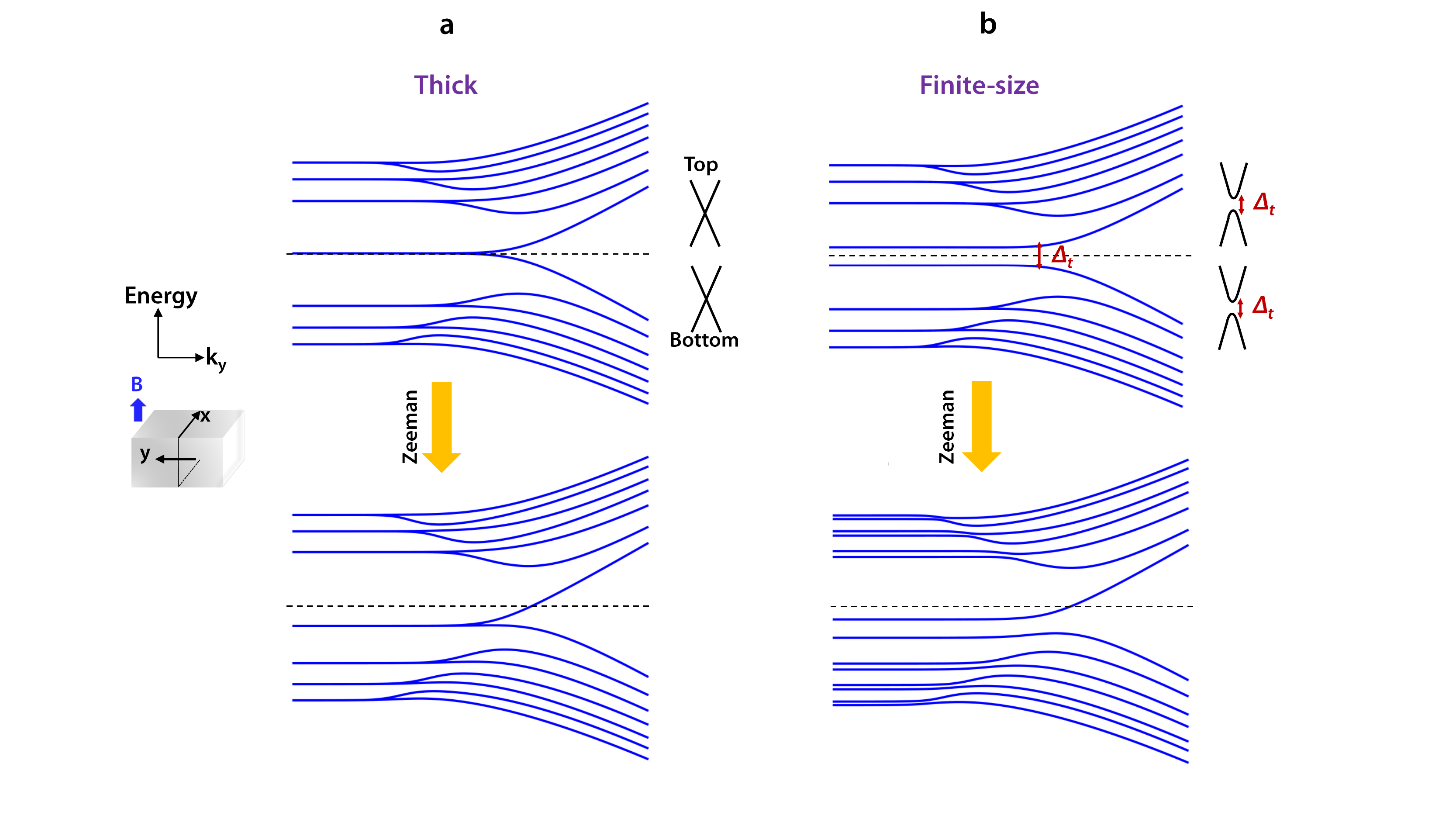}
\caption{\label{fig:edgetunnel} Surface LLs and edge modes on the side surfaces of a) a thick TI sample without (top panel) and with (bottom panel) Zeeman field and b) a hybridized TI with a finite-size tunneling gap of $\Delta_{t}$ which lifts the degeneracy of the zeroth LLs. The surface LL energies are given by (\ref{eq:surf_LL_evals}) and (\ref{eq:landaulevels}).
Although, in practice this gap is smeared by disorder and could be effectively zero, applying sufficiently large magnetic field could resolve this gap to the clean-limit (see discussion below Fig. \ref{fig:phasediag}).  Higher magnetic fields could increase this gap further through band structure effects explained below Eq.(\ref{eq:zeromode_B}). The Zeeman energy shift of LLs is also shown in the bottom panel of b.  }
\end{figure}

\newpage

\begin{figure}[H]
\centering
\includegraphics[scale=0.35]{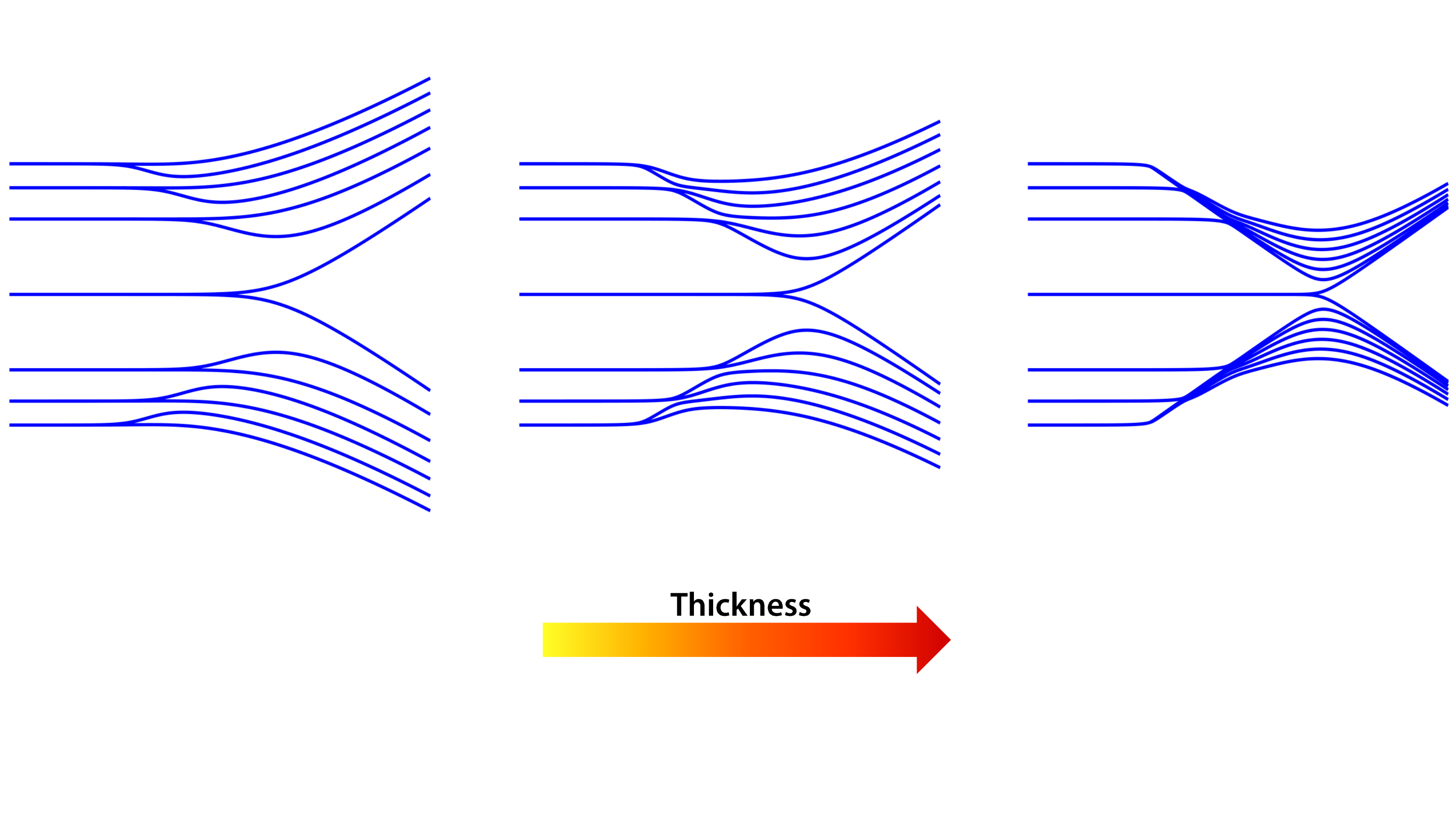}
\caption{\label{fig:edgethickness} Evolution of surface LLs along with edge modes on the side surfaces in a TI system as a function of thickness (thinner to thicker from left to right). The spectrum of edge modes on the side surfaces approaches a 2D Dirac cone spectrum  as the sample gets thicker.}
\end{figure}

\newpage
\begin{figure}[H]
\centering
\includegraphics[scale=0.15]{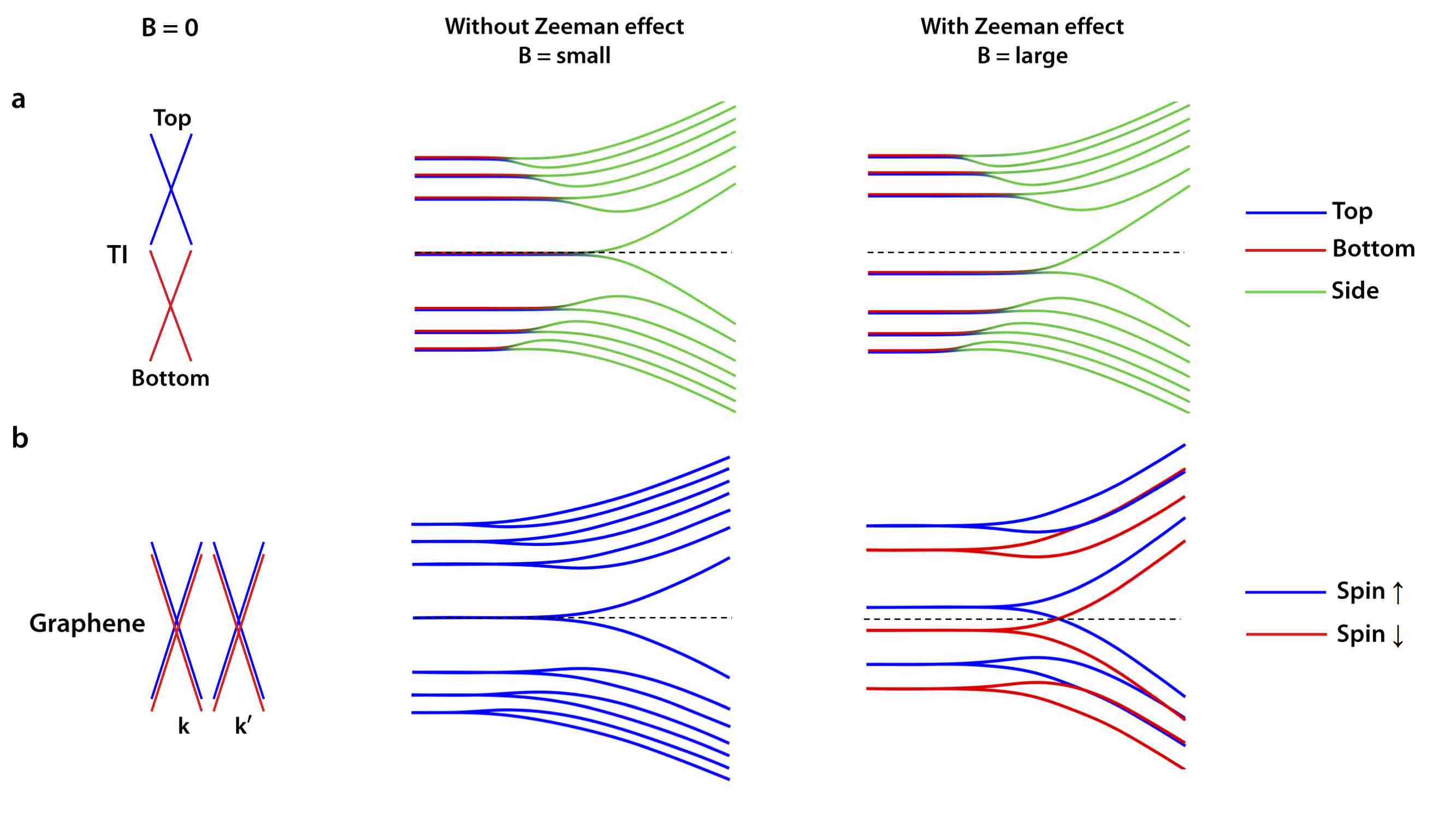}\caption{\label{fig:TIvsgraphene}Schematic of energy spectrum of a) TI thin film and b) graphene in the absence of magnetic field (the first column) and in the presence of magnetic field with (the second column) and without (the third column) considering the Zeeman-coupling. The zero energy is marked by dashed lines. Because of spin-momentum locking in TI surface Hamiltonian, the Zeeman field does not lift the LL degeneracies, instead, it shifts all LLs.  Particularly, the Zeeman term moves the top and bottom LLs up or down depending on magnetic field direction. The energy shift is more pronounced for the zeroth LLs and becomes smaller for higher LLs (c.f.~Eq.(\ref{eq:landaulevels})).  In contrast,  in graphene the Zeeman field lifts the spin degeneracy and  gaps the zeroth LL. The band crossing of the counter-propagating edge modes in graphene can be gapped by impurity scattering, effectively driving the system to an insulating regime~\cite{checkelsky,Geim,Giesbers}.}
\end{figure}

\newpage

\bibliography{refs}

\end{document}